\documentclass[%
 reprint,
nofootinbib,
nobibnotes,
bibnotes,
 amsmath,amssymb,
 aps,
]{revtex4-1}
\usepackage{MnSymbol}
\usepackage{natbib} 
\usepackage{xcolor}
\usepackage{soul}
\usepackage{wasysym} 
\usepackage{graphicx}
\usepackage{dcolumn}
\usepackage{bm}

\begin{document}

\preprint{Draft}


\title{An exact $N$-strain epidemic model using bond percolation}

\author{Peter Mann}
\email{pm78@st-andrews.ac.uk}
\author{V. Anne Smith}%
\author{John B.O. Mitchell}
\author{Simon Dobson}
\affiliation{School of Computer Science, University of St Andrews, St Andrews, Fife KY16 9SX, United Kingdom }
\affiliation{EaStCHEM School of Chemistry, University of St Andrews, St Andrews, Fife KY16 9ST, United Kingdom }
\affiliation{School of Biology, University of St Andrews, St Andrews, Fife KY16 9TH, United Kingdom }

\date{20 July, 2022}

\begin{abstract}
    In this paper we examine the emergent structures of random networks that have undergone bond percolation an arbitrary, but finite, number of times. We define two types of sequential branching processes: a competitive branching process - in which each iteration performs bond percolation on the residual graph (RG) resulting from previous generations; and, a collaborative branching process - where percolation is performed on the giant connected component (GCC) instead. We investigate the behaviour of these models, including the expected size of the GCC for a given generation, the critical percolation probability and other topological properties of the resulting graph structures using the analytically exact method of generating functions. We explore this model for Erd\H{o}s-Renyi and scale free random graphs. This model can be interpreted as a seasonal $N$-strain model of disease spreading. 
\end{abstract}

\pacs{Valid PACS appear here}
\maketitle

\section{Introduction}
\label{sec:intro}

\begin{figure*}[ht!]
  \begin{center}
    \includegraphics[width=0.99\textwidth]{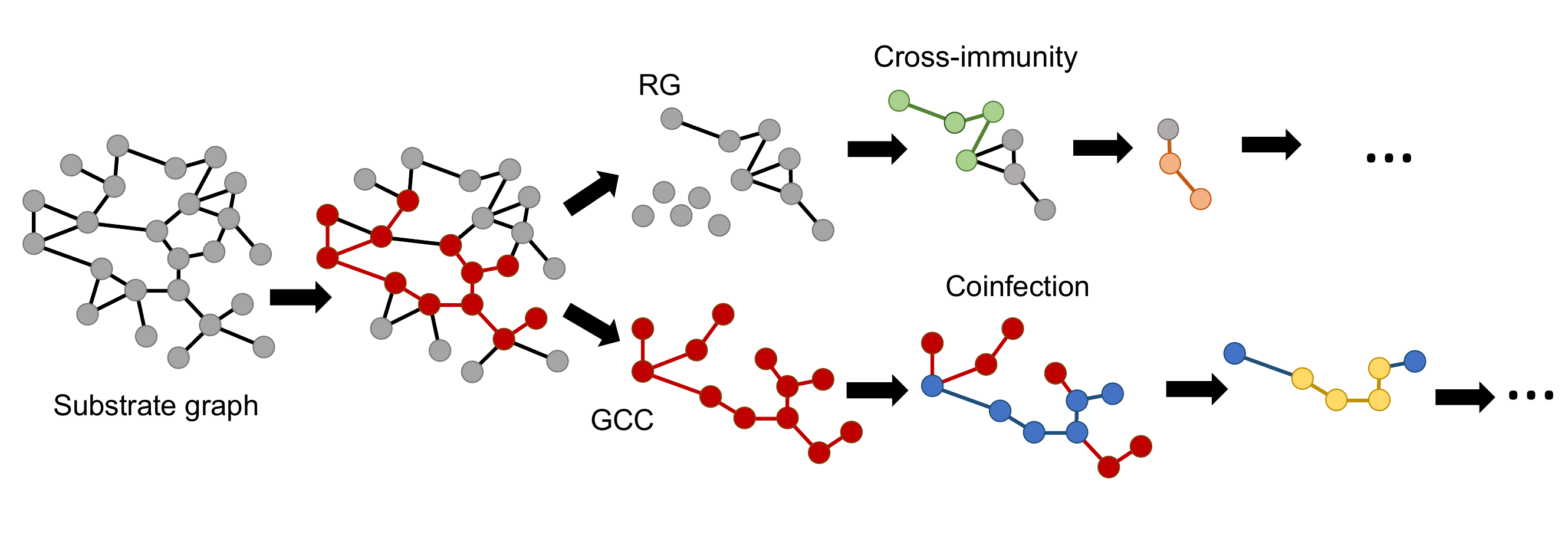}
  \end{center}
  \caption{A substrate network (left) is percolated to yield a GCC (red) and an RG. In the competative branching process, the RG is percolated again to create another GCC (green) and so on. This is a cross-immunity model. The collaborative branching process instead percolates the GCC of each generation and constitutes a  model of complete coinfection.   
  }
  \label{fig:nicefigurefull}
\end{figure*}

A network is a collection of nodes connected by edges. Bond percolation on complex networks is a widely studied binary-state stochastic process. In this model, the edges of a graph are iterated and said to be \textit{occupied} with probability $T$ and remain \textit{unoccupied} with probability $1-T$. When $T$ is small, the network is comprised of many small isolated components of occupied edges. At some critical value, $T_c$, the small components connect together to form a macroscopic giant connected component (GCC), exhibiting a second-order phase transition. As $T\rightarrow 1$ the GCC occupies an increasing fraction of the network. Nodes not contained within the GCC are said to be in the residual graph (RG) of the percolation process, see Fig \ref{fig:nicefigurefull}. 
The size of the GCC following bond percolation, as well as the value of the critical bond occupation probability, have an equivalence to the absorbing state of the susceptible, infected, removed (SIR) epidemic process \cite{Newman2002SpreadOE} when a node's infectious period is drawn from a single-valued distribution \cite{PhysRevE.76.036113}. In the SIR model, nodes are either susceptible to infection from their infected neighbours; infectious, or are in the removed state. Transmission of infection occurs along edges that connect infected nodes to their susceptible neighbours. Once infected, a node remains infectious for a fixed period \cite{PhysRevE.76.010101,PhysRevE.76.036113}, $\tau$, before recovering to the $R$ state. Hence, the absorbing equilibrium of the model is a static binary state that is composed of susceptible nodes that did not contract the disease or removed nodes that did. The size of the GCC following bond percolation is equivalent to the largest outbreak size of the SIR model.  

There has been previous work on extending the percolation-SIR equivalence to additional, temporally separate, strains of an epidemic for random graphs \cite{newman_2005,PhysRevE.84.036106,newman_ferrario_2013,PhysRevE.81.036118,BANSAL2012176,PhysRevE.103.042307,PhysRevE.103.062308,cai_chen_ghanbarnejad_grassberger_2015,bansal_pourbohloul_hupert_grenfell_meyers_2010}. Such models can be thought of as a model of seasonal diseases; each pathogen running its course through a population before the next season's strain spreads. Within this context, the nature of how the second strain interacts with nodes that have been previously infected is very important to the resultant disease dynamics. Newman \cite{newman_2005,PhysRevE.84.036106} studied two strains that compete for hosts by not allowing nodes in the removed state of strain 1 to become infected by strain 2. This model is an example of perfect cross-immunity. Mann \textit{et al.} extended this work to study the role of contact clustering and modularity on the spread of the second pathogen \cite{PhysRevE.103.062308}. Conversely, Newman and Ferrario \cite{newman_ferrario_2013} studied the opposite scenario in which only those nodes that have been infected by strain 1 can be later infected by strain 2. This is an example of a perfect coinfection model and the extension to the clustered case was also studied \cite{PhysRevE.103.042307}. In each case, the second strain is a separate SIR process that either spreads exclusively on the RG (cross-immune) or the GCC (coinfection) of the first percolation process, see Fig \ref{fig:nicefigurefull}. Both processes are abstractions of the spread of real diseases; since, the strict requirements of perfect cross-immunity and perfect coinfection are the two limits of a spectrum of partial interactions. Partial cross-immunity and partial coinfection, passing through a point of complete decoupling is known as a partial immunity model and has recently been studied using bond percolation \cite{PhysRevE.104.024303} for the case of two sequential percolation processes (diseases). 

Whilst these models are of clear interest to an epidemiology readership, the implications for the understanding of network structure cannot be understated. In particular, the cross-immune model provides us with a model of the RG of a network following bond percolation; similarly, the coinfection model yields insight into the structure of the GCC.

In this paper we extend the percolation models of \cite{newman_2005,PhysRevE.84.036106,newman_ferrario_2013} from two sequential strains to an arbitrary, but finite, number $N\in \mathbb N$ of sequential strains. We study the topological properties of the graph structures that arise; quantifying the expected outbreak sizes for each disease and the critical points of the model as well as their degree distributions and cumulative degree distributions of the graph structures that emerge.

\section{Generating functions}
\label{sec:genfunc}

The study of a bond percolation process over a network using generating functions was pioneered by Newman, Strogatz and Watts around 20 years ago \cite{newman_strogatz_watts_2001,Newman2002SpreadOE,newman_2019}. In this section we will review the key components of the generating function formulation that our $N$-strain model will use. Generating functions are infinite series; in the context of network science the summation is over the degrees of the nodes, $k$, whilst the coefficients are the values of the degree distribution, $p_k$, the probability of choosing a degree $k$ node at random. This is generated by 
\begin{equation}
    G_0(x) = \sum_{k=0}^\infty p_kx^k
\end{equation}
The coefficients of a generating function can be recovered by repeated differentiation
\begin{equation}
    p_k=\frac{1}{k!}\frac{d^kG_0}{dx^k}\bigg|_{x=0}
\end{equation}
It is commonplace to evaluate the derivatives by numerical contour integration using the Cauchy formula
\begin{equation}
   p_k= \frac{1}{2\pi i}\oint \frac{G_0(z)}{z^{k+1}}dz\label{eq:degreedist}
\end{equation}
We can obtain the cumulative distribution function (CDF), $p_{k<\nu}=\sum\limits^\nu_{k=0}p_k$, defined as the probability that a randomly chosen node has a degree less than or equal to $\nu$, directly from $G_0(x)$. To achieve this, it is convenient to instead calculate one minus the probability of a node having a degree \textit{greater or equal to} $\nu$, $p_{k\geq \nu}$, which is called the survival function \cite{wilf_1990}. The latter distribution is generated by 
\begin{equation}
    G_{0,\nu}(x) = \frac{G_0(x) - p_0 - \dots - p_{\nu - 1}x^{\nu-1}}{x^\nu}\label{eq:cumulative}
\end{equation}
The values of $p_k$ can be obtained from Eq \ref{eq:degreedist} and evaluating $G_{0,\nu}(x)$ at $x=1$ yields $p_{k\geq \nu}$. 

The probability of choosing an edge at random from the network, following it randomly to a particular end to reach a node of degree $k$ is generated by 
\begin{equation}
    G_1(x) = \frac{G_0'(x)}{G_0'(1)}\label{eq:g1g1}
\end{equation}
where $G_0'(1) = \langle k\rangle$ is the average degree of the network. 


For example, one of the simplest degree distributions occurs when the degrees are binomially distributed; this is the Erd\H{o}s-Renyi degree distribution. In this model, the probability, $\theta=\langle k \rangle/V$, where $V$ is the number of nodes, is the probability that an edge connects any two nodes, for all pairs of nodes in the network. In this instance, we find 
\begin{equation}
    G_0(x) = \sum_{k=0}^V\binom{V}{k}\theta^k(1-\theta)^{V-k}x^k
\end{equation}
which, in the limit of large $V$, is equal to $ G_0(x) = e^{\langle k\rangle (x-1)}$ It must be noted, in this special case, that $G_1(x) = G_0(x)$ which simplifies the model significantly. 

\subsection{Cavity method}

The cavity method is a statistical technique that is commonly used when dealing with generating functions. The method considers the local environment of a particular node chosen from the network at random (the focal node) once a (percolation) process has reached equilibrium. We suppose that the focal node's degree is $k$. We then calculate the probability associated with all configurations of the node's neighbours before averaging over the probability that the node we chose did indeed have degree $k$. As a motivating example of this method, consider a graph whose nodes are in one of two states, $A$ or $B$, after a binary-state process has reached a static absorbing state. The focal node will have $k_A\leq k$ neighbours that are in state-$A$ and $k_B = k-k_A$ neighbours are in the mutually exclusive state-$B$. Assuming that the neighbour-state probabilities are independent and identically distributed (iid) along each edge the sum over all configurations of the probability of configuration $k_A$ and $k_B$ given $k$ is
\begin{equation}
  g^k = \sum^k_{k_A=0}\binom{k}{k_A}(\pi_Ax)^{k_A}(1-\pi_A)^{k-k_A}
\end{equation}
Each combination of $k_A$ and $k_B$ is then wrapped into the $k$th term of a generating function as we average over the probability that the focal node we chose had degree $k$, $\sum_kp_kg^k$. This method is key to the analytical description we present in this paper. We will continually examine the permissible neighbours that connect to our focal node and their associated probabilities of occurring along each iid edge. As the number of generations of the disease process increases, the number of permissible neighbour states grows too; and so the local environment that an embedded node experiences evolves into a rich and complex landscape. 

\section{Competitive Branching process}
\label{sec:type1}

In this section, we define the $i$th competitive branching process as successive bond percolations occurring on the RG created by the $i-1$ previous processes for $i=1,\dots,N$. This process has been studied previously using generating functions by Newman and Karrer when $N=2$ \cite{newman_2005,PhysRevE.84.036106}. The structure of the RG has also been studied for clustered and modular networks \cite{PhysRevE.103.062308}. From a network science perspective, this model allows us to study the structure of the RG of sequential bond percolation processes. In particular, we observe how those sequential processes fracture the RG into isolated components and study the phase behaviour associated with the sudden inability of the RG to support a GCC. Within the context of the SIR equivalence, this model considers the behaviour of $N$ seasonal strains of a disease (or separate diseases) that confer complete cross immunity to all subsequent pathogens. The model allows us to study the expected outbreak size of each generation and the point of natural burn-out due to the shrinking of the susceptible sub-population. 

\subsection{Outbreak size}

To the $i$th process, $i=1,\dots,N$, we assign a bond occupation probability $T_i$ and aim to calculate the probability that a randomly chosen node does not belong to the largest percolated component of that generation. From this, we can find the mutually exclusive probability that a node does belong to the $i$th GCC, $\mathcal A_i$. To do this, we define the probability that a neighbour of our randomly selected node is not part of the $i$th GCC, $u_i$, given that it does not belong to any of the previous percolated components. Under the SIR equivalence, $T_i$ is the transmissibility of the $i$th strain and $u_i$ is the probability that a neighbour is not thus far infected.

There are two ways in which an edge emanating from the focal node can fail to connect it to the GCC: firstly, the neighbour could itself be unconnected, the probability of which by definition is $u_i$. Secondly, the neighbour could be connected, $(1-u_i)$, but the bond is unoccupied $(1-T_i)$. Therefore, the probability, $\bar g_i$, that an edge fails to connect the focal node to the $i$th GCC given that the neighbour does not belong to any other GCC is
\begin{equation}
   \bar g_i(T_i,u_i\mid \text{RG}) = u_i + (1-u_i)(1-T_i)\label{eq:gbar}
\end{equation}

The total probability that a neighbour belongs to the RG  of the $i$th percolation can then be found through a set of recursive functions, $g_i$ that describe the probability that each iterative percolation failed to occupy this edge as
\begin{equation}
    g_{i}(\bm T,\bm u) = u_{i-1}\bar g_i + (1-u_{i-1})(1-T_{i-1})\label{eq:gnorm}
\end{equation}
with $u_0=1$. A hierarchy of self-consistent equations can be written to sequentially solve for each $u_i$ value
\begin{equation}
    u_{i} = \frac{G_1(g_i)}{\prod_j u_j},\qquad j = 1,\dots ,i-1 \label{eq:u_cross}
\end{equation}
The size of the $i$th GCC (epidemic) is then found by 
\begin{equation}
    \mathcal A_i = \prod_{j=1}^{i-1}G_0(g_j) - G_0(g_i)\label{eq:maintype1}
\end{equation}
The total infected fraction of the network is given by $ \mathcal A = \sum_i\mathcal A_i$.

As an example of Eq \ref{eq:maintype1}, we can obtain the expected outbreak size of the first epidemic, $i=1$, from this system as $\mathcal A_1=1-G_0(g_1)$ where $u_1=G_1(g_1)$ and $g_1=u_1+(1-u_1)(1-T_1)$, \cite{Newman2002SpreadOE}. In the case that $i=2$ \cite{newman_2005}, we have $\mathcal A_2=G_0(g_1)-G_0(g_2)$, where $u_2 = G_1(g_2)/u_1$ and 
\begin{align}
g_2=&u_1(u_2+(1-u_2)(1-T_2))\nonumber\\
&+(1-u_1)(1-T_1)
\end{align}
Similarly, for $i=3$ we have $\mathcal A_3=G_0(g_1)G_0(g_2)-G_0(g_3)$ with  $u_3=G_1(g_3)/(u_1\cdot u_2)$ and 
\begin{align}
g_3=&u_1(u_2(u_3+(1-u_3)(1-T_3))\nonumber\\
&+(1-u_2)(1-T_2))+(1-u_1)(1-T_1)
\end{align}
With these examples, it is hopefully clear how to write further generations of the competitive percolation process. 

Given this, we can derive the conditions under which $\mathcal A_i$ is maximised given that the previous strains have run their course of the network. To see this, consider that each expression $G_1(g_i)$ has at most two roots for $u_i$: the trivial case $u_i=1$, corresponding to no GCC ($\mathcal A_i=0$); and, a non-zero root in the unit interval that indicates that a finite fraction of the network is occupied by this strain ($\mathcal A_i>\epsilon$ for some non-zero fraction $\epsilon$). Since the coefficients of $G_1(g_i)$ are non-negative, so too are its derivatives (for $u_i\geq 0$). Hence, $G_1(g_i)$ is in general positive and convex. Given that each $g_i$ is convex in $u_i$ for $T_i\in [0,1]$, then, by Jensen's inequality \cite{jensen_1906}, there exists a unique root $T_i^*$ that minimises $g(u_i)$. Thus, the best case scenario for the $i$th strain to maximise $\mathcal A_i$ is to have $g_j= 1$ and $g_i= 0$ for $j=1,\dots,i-1$. This is obtained by all previous strains adopting the minimising transmissibility of $T_j^*=0$ and the $i$th strain adopting $T_i=1$.

\subsection{Invasion threshold}

The $\mathcal R_0$ value is defined as the  number of new infections caused by an infected individual. When $\mathcal R_0<1$, the epidemic fails to infect a significant portion of the network; the GCC comprises $\mathcal O(1)$ nodes. When $\mathcal R_0=1$, the probability that an epidemic infects a macroscopic fraction of the network, $\mathcal O(V)$, is non-zero, where $V$ is the number of nodes in the network. This point is also the critical point in bond percolation that marks the smallest value of $T$ that can form a GCC.   
Within our model, there is an $\mathcal R_{0,i}$ value and a critical transmissibility for each strain. This critical point is a function of both the network topology and the transmissibilities of the previous strains. If the transmissibility of a particular strain is below its critical threshold, it only infects $\mathcal O(1)$ nodes before it burns out. Therefore, in the following analysis, we assume that the transmissibility of each strain is greater than its minimum threshold.  

The critical point for the $i$th percolation can be found by applying linear stability analysis on the system given in Eq \ref{eq:u_cross} around the fixed point $u_i=1$. This is the point at which the fixed point in $u_i$ bifurcates into two solutions and $\mathcal A_i$ becomes non-zero. Performing a Taylor expansion around $\epsilon = 1-u_i$ we find the following condition
\begin{align}
    \epsilon_i = 1 - \frac{1}{\prod_j u_j}\frac{\partial G_1(g_i)}{\partial g_i}\frac{\partial g_i}{\partial u_i}\epsilon_i\bigg|_{u_i=1} + (\mathcal O^2)\label{eq:1MR}
\end{align}
where the derivatives are given by 
\begin{equation}
    \frac{\partial g_i}{\partial u_i} = T_i\prod_{j=1}^{i-1}u_j 
\end{equation}
When evaluated at $u_i=1$, $G_1'(g_i)$ becomes $G_1'(g_{i-1})$ from Eqs \ref{eq:gbar} and \ref{eq:gnorm}. The critical transmissibility is found to be 
\begin{equation}
    T_{i,c} = \frac{1}{G_1'(g_{i-1})}\label{eq:criticalTtype1}
\end{equation}
Thus, the minimum transmissibility required for each strain to create an epidemic is a function of the network topology and the transmissibilities of the preceding strains.  Given that the coefficients of $G_1'(x)$ are non-negative and therefore monotonically increasing on the positive real line (within its radius of convergence), and that $g_j\in [0,1]$ (since it is a probability) and because $T_j,u_j\leq 1$, then it follows that $G_1'(g_i)\leq G_1'(g_{i-1})$ $\forall i$. This indicates (from Eq \ref{eq:1MR}) that $T_{i,c}\geq T_{i-1,c}$. In other words, the epidemic threshold of each strain increases with each generation. This is an intuitive result, since, as each strain passes through the network, nodes with higher degree are preferentially embedded into the GCC of that strain. Therefore, the RG is increasingly comprised of lower degree nodes as it fractures with each iteration of the percolation. 

\begin{figure}[ht!]
  \begin{center}
    \includegraphics[width=0.5\textwidth]{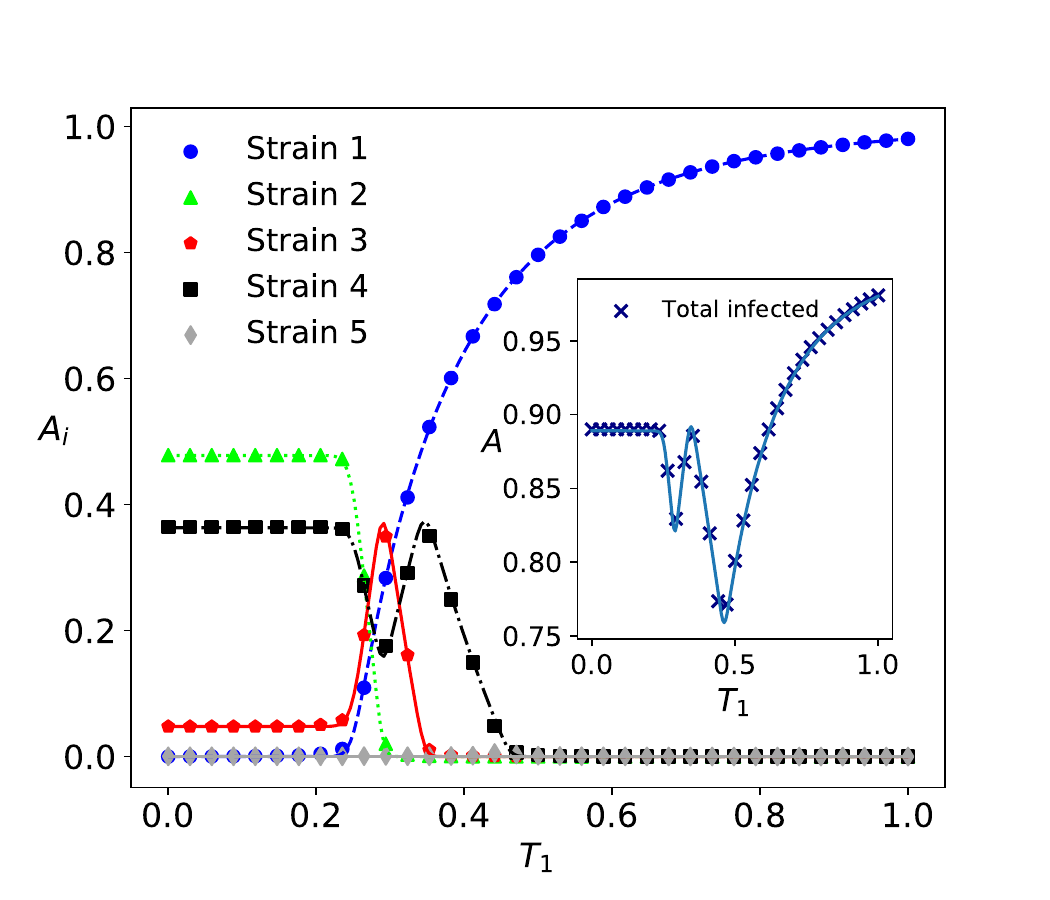}
  \end{center}
  \caption{The outbreak fractions for five generations of the competitive branching process as a function of $T_1$. Solid lines are the theoretical results from Eq \ref{eq:maintype1} whilst scatter points are the average of 50 repetitions of bond percolation over a network with $V=35000$ nodes. The inset shows the total number of infected nodes. 
  }
  \label{fig:iterated_type1}
\end{figure}

\begin{figure}[ht!]
  \begin{center}
    \includegraphics[width=0.40\textwidth]{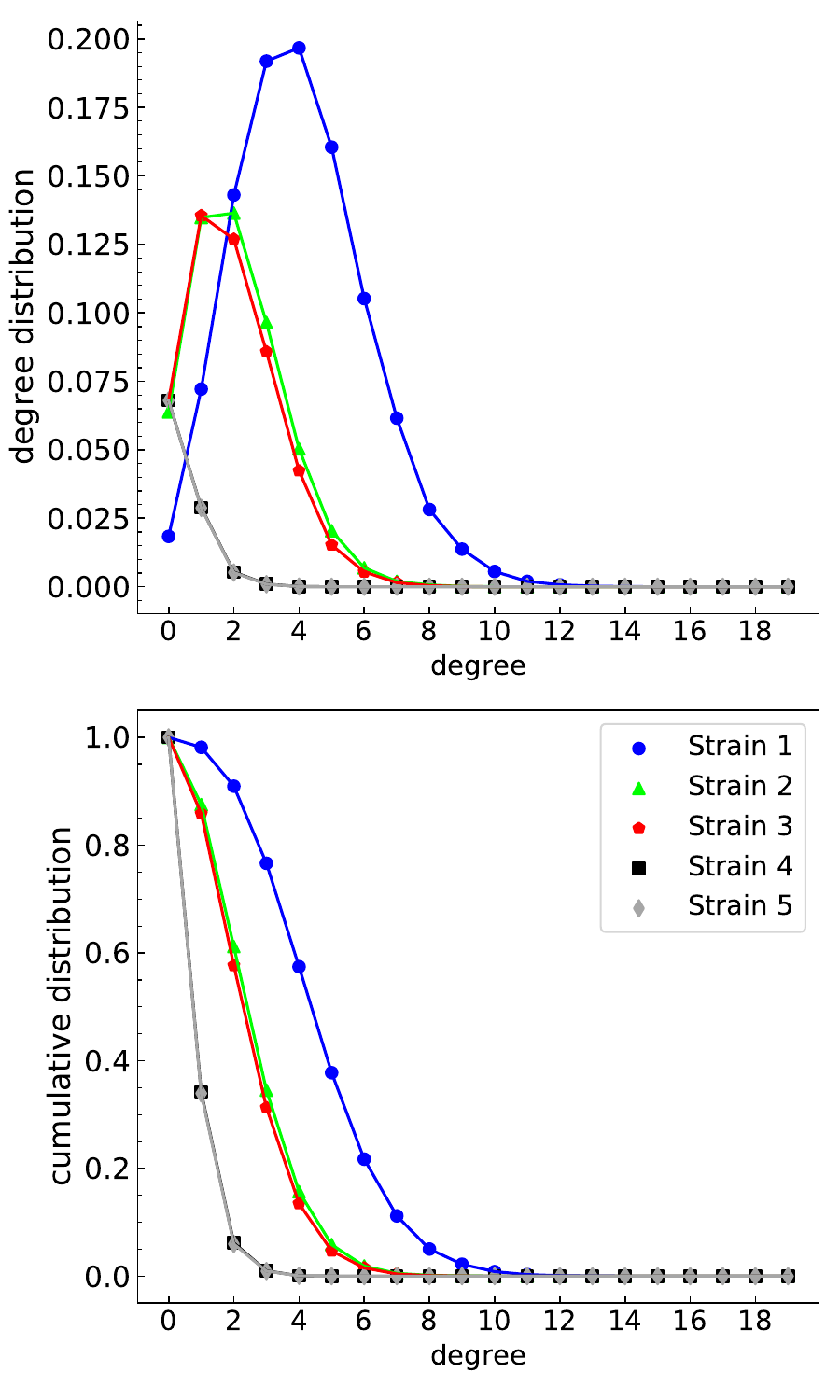}
  \end{center}
  \caption{(top) The degree distribution of the RG created once each strain from Fig \ref{fig:iterated_type1} has spread over the network at $T_1=0.0$; and, the corresponding cumulative degree distribution (bottom). Scatter points are the average of 50 repetitions of $N=35000$ Erd\H{o}s-Renyi networks with $\langle k \rangle =4$. Curves are the result of Eqs \ref{eq:degreedist} and  \ref{eq:cumulative} acting on the generating functions for the residual graphs of each generation. These plots show how the RG becomes increasingly fractured with each percolation.
  }
  \label{fig:iterated_type1_cums}
\end{figure}

Following \cite{PhysRevE.84.036106} we can also prove a stronger condition on the minimum bond occupation probability that a subsequent strain must have in order to exhibit an epidemic on the network. It happens that each generation of the disease must have an increasingly higher transmissibility than the last in order to infect $\mathcal O(V)$ nodes in the RG. To see this, we note that $g_i(\bm T,\bm u)$ is the probability that an edge fails to connect a node to the GCC of the $i$th epidemic and that this probability can only decrease or stay constant as $T_i$ increases; this implies that $d T_i/dg_i\leq 0$. Inverting this quantity such that $T_i=T_i(\bm u,g_i,\bm T\backslash\{T_i\})$ where the notation $\bm S \backslash \{s\}$ excludes element $s$ from set $\bm S$, and performing the derivative we have an expression that involves $T_i$ and $ G_1'(g_{i-1})$. This can then be isolated and it can be shown that $T_{i,c}\geq T_{i-1}$ $\forall i\in [1,N]$. For example, the critical point of strain 2 is known \cite{newman_2005, PhysRevE.84.036106} to be greater than the transmissibility of strain 1, $T_1$. The critical point of strain 3 is given by $T_{3,c}=1/G_1'(g_2)$ from Eq \ref{eq:1MR} which, through the above prescription satisfies 
\begin{equation}
    T_{3,c} \geq \frac{g_1-g_2}{1-u_2} =T_2
\end{equation}
This logic can be applied to all adjacent strains to create an ordered set of critical transmissibilities $\{T_{1,c}\leq T_{2,c},\dots,T_{n,c}\}$. This indicates that transmissibility must evolve to increase in order for a given strain to create an epidemic in the presence of others. 

The coexistence threshold $T_x$ was defined in \cite{newman_2005,PhysRevE.84.036106} for two pathogens and marks an additional phase transition in the model. For $N=2$ it is the largest value of $T_1$ that still allows the RG to retain sufficient connectivity to support its own GCC for future strains. For instance, when $T_1>T_x$, the RG fails to be globally connected and strain 2 fails to infect $\mathcal O(V)$ nodes even if $T_2>T_{2,c}$. For our purpose we extend the definition of the coexistence threshold, $T_{i,x}$,  in the context of $N$ sequential strains to be the largest transmissibility of strain-$i$ that allows the RG to support a GCC for future generations, assuming that they are sufficiently transmissible. Thus, $T_{i,x}$ is a function of the bond occupancy probabilities of all previous percolations, $T_{i,x}= T_{i,x}(\bm u,\bm T\backslash\{T_i\})$. As for $N=2$ \cite{newman_2005,PhysRevE.84.036106}, Eq \ref{eq:1MR} implicitly defines the coexistence threshold of the $i$th strain and we find that $T_{i,x}$ is the value of $T_i$ for which $G_1'(g_{i})=1$. For instance, for an Erd\H{o}s-Renyi degree distribution this condition becomes \begin{equation}
    \frac{1}{\langle k\rangle}\ln \left[u_i\prod_{j=1}^{i-1}u_j\right] = g_i-1\label{eq:coexistence}
\end{equation}
from which we can solve for $T_{i,c}$ by inverting $g_i$. For $i=1$ we have 
\begin{equation}
    T_{1,x} = \frac{\ln(u_1)}{\langle k \rangle} \frac{1}{(u_1-1)}
\end{equation}
In Fig \ref{fig:iterated_type1} we plot $\mathcal A_i$ for $N=5$ against $T_1\in [0,1]$ and $T_2=0.35$, $T_3=0.5$, $T_4=1.0$ and $T_5 = 1.0$ for a Erd\H{o}s-Renyi random graph with mean degree $\langle k \rangle=4$ and $V=35000$ nodes. We observe excellent agreement between experimental bond percolation (scatter points) and the analytical results of Eq \ref{eq:maintype1} (plotted lines). Below the epidemic threshold of the first strain, $T_1<T_{1,c}$, strain 1 does not exhibit a GCC. Hence, the RG is large enough to enable the subsequent strains to form their own GCCs, each consuming more of the available space. With $T_4=1$, the last edges in the RG are occupied and, despite a supercritical $T_5$, we have $\mathcal A_5=0$. Strain-4 is bimodal, exhibiting two turning points as a function of $T_1$. This is because, at the first turning point, the transmissibility of the previous strains is sufficient to form their own large GCCs in the RG; however, as $T_1$ increases, strains 2 and 3 fall below their critical thresholds, allowing strain-4 to consume the available sites into its own GCC. The outbreak size of strain-4 then falls to zero through a final turning point as the transmissibility of strain-1 is increased beyond the coexistence threshold. The inset figure shows the total fraction of the network that has become infected, $\mathcal A$ versus $T_1$. Against intuition, the largest fraction of the network that is occupied by (any) disease, is not constant. To see this, we understand that the early generations of the disease consume the high degree sites. As these become embedded within the GCC, those nodes they connect to can become isolated and thus, cannot be incorporated in subsequent GCCs (see Fig \ref{fig:iterated_type1_cums}). The effect of this is prominent at the onset of the GCC of strain 1 and leads to a local minimum in the total infecteds. Therefore, a disease of low transmissibility early on not only consumes nodes into its own GCC, but also reduces the accessible sites by removing the infection pathways. As the transmissibility of the initial pathogen increases, this effect is reduced and the total infected fraction of the network increases to $\mathcal A \approx 0.9$. The global minimum at $T_1\approx 0.45$ coincides with the coexistence threshold for strain 1; we observe the inability of subsequent strains to create their own epidemic. At this point, strain 1 is sufficiently transmissible to fracture the RG such that it can no longer support a GCC  for the other generations of the disease. Beyond this point the total infected fraction follows the number of infected nodes of strain 1. 

\begin{figure}[ht!]
  \begin{center}
    \includegraphics[width=0.40\textwidth]{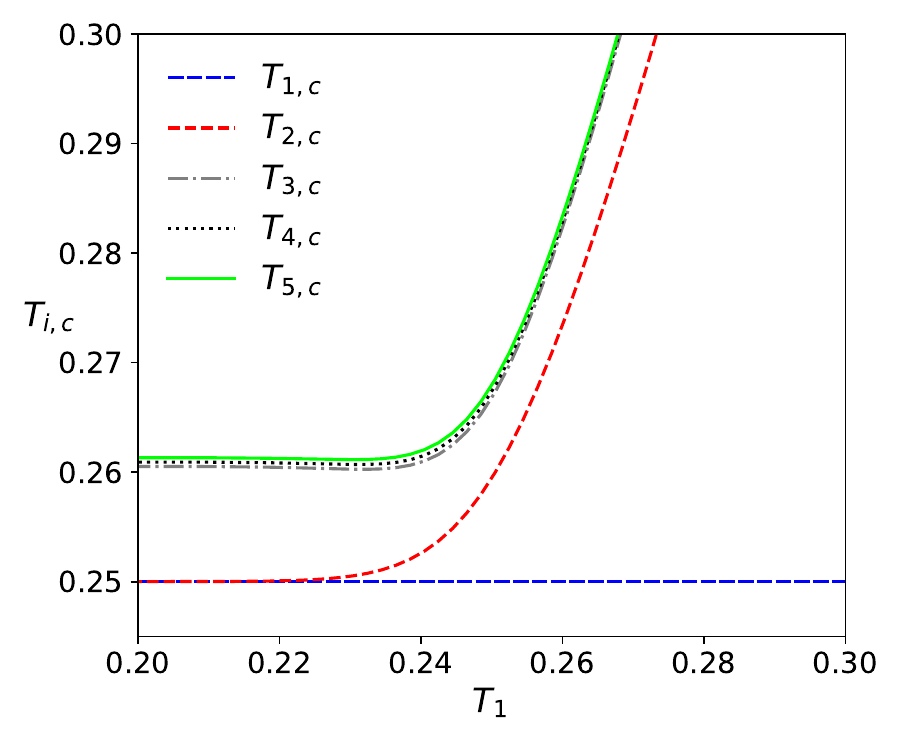}
  \end{center}
  \caption{
  The critical points of the first five strains of the competitive percolation process as a function of strain 1 transmissibility from Eq \ref{eq:1MR}. The transmissibility of each strain is set to the critical point plus a small parameter such that $T_i=T_{i,c}+\delta$ with $\delta = 1\times 10^{-4}$. The networks are Erd\H{o}s-Renyi graphs with mean degree $\langle k \rangle = 4$. 
  }
  \label{fig:iterated_type1_crit5}
\end{figure}

In a second experiment, we examine the critical points of the first five strains (see Fig \ref{fig:iterated_type1_crit5}) versus disease-1 transmissibility. To do this, we set the transmissibility of each strain to its critical point, $T_{i,c}$, perturbed by a small parameter, $\delta$, such that $T_i=T_{i,c}+\delta$ with $\delta = 1\times 10^{-4}$. From the logic above, this scenario represents the best-case for the $i$th strain to form a GCC subject to the constraint that each previous disease did indeed form an epidemic. When $T_1< T_{1,c}$ the epidemic threshold of the second strain is equal to $1/\langle k \rangle$. As the first process forms a GCC, the substrate RG available to the second percolation process fractures, resulting in an increase in the critical point for the second strain. When there are one or more GCCs in the network, the critical points of subsequent strains are increased. This effect is most significant at the formation of the first GCC, indicating that the RG suddenly `fails' to be well connected. In the presence of multiple GCCs, the epidemic threshold for subsequent strains, $(i>2)$, is only slightly higher than the last. 

\begin{figure}[ht!]
  \begin{center}
    \includegraphics[width=0.52\textwidth]{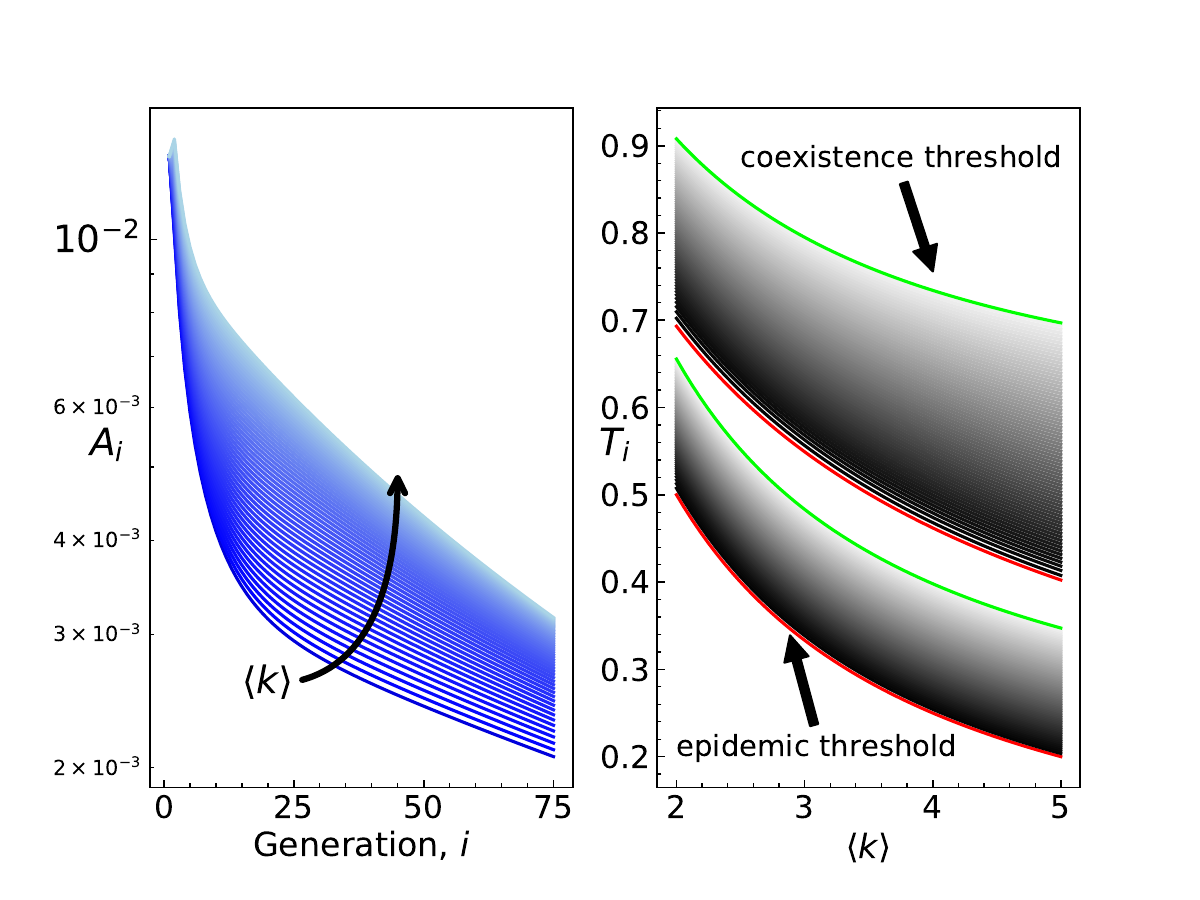}
  \end{center}
  \caption{
(Left) Theoretical results for the outbreak size of the first $N=75$ generations of the competitive percolation for Erd\H{o}s-Renyi graphs with varying mean degree $\langle k \rangle$. (Right) The evolution of the epidemic threshold and the coexistence threshold for each generation, $i$. The lines for each subsequent strain are plotted from dark to light greyscale whilst the curves belonging to strain 1 have been coloured red, whilst those of strain $N=75$ are hilighted in green.     
  }
  \label{fig:iterated_type1_many}
\end{figure}

In Fig \ref{fig:iterated_type1_many} we examine the outbreak size predicted by the model for the first $N=75$ strains on Erd\H{o}s-Renyi graphs as a function of mean degree; finding that the expected outbreak size increases with $\langle k \rangle$. In similar fashion to before, the transmissibility of each strain is set equal to $T_i=T_{i,c}+\delta$. In the left figure, we see $\mathcal A_i$ monotonically decreasing with generation index $i$. Thus, whilst the transmissibility evolves to increase, the number of infected hosts is expected to decrease with each season of the disease. We numerically solve Eq \ref{eq:coexistence} to obtain the coexistence threshold and analytically solve Eq \ref{eq:1MR} to find the epidemic threshold of the $i$th generation (Fig \ref{fig:iterated_type1_many} right). The two curves for the first strain are plotted in red, whilst subsequent strains are lighter greyscale until the final generation, which has been highlighted green. We observe that both critical points of the model occur at lower transmissibilities with increasing $\langle k \rangle$ as supported by \cite{newman_2005} for $N=2$. The coexistence threshold broadens to a larger extent than the the epidemic threshold with increasing $i$. In other words, $|T_{i-1,x}-T_{i-1,c}|\leq |T_{i,x}-T_{i,c}|$ indicating that strain coexistence occupies an increasingly larger area of the model's phase space with greater $i$. 

\section{Collaborative branching process}
\label{sec:type2}

In this section we define the $i$th generation of a collaborative branching process as a bond percolation process occurring on the GCC that is created by the $i-1$ previous processes for $i=1,\dots,N$. We impose the strict requirement that only nodes in the GCC created by all of the previous generations are included at the $i$th generation see Fig \ref{fig:nicefigurefull}. This process has been studied previously by Newman and Ferrario when $N=2$ \cite{newman_ferrario_2013} as well as for clustered and modular networks \cite{PhysRevE.103.042307}. Within the context of the SIR equivalence, this model studies the ability of the $i$th disease to become an epidemic given that coinfection with all other $i-1$ strains is a prerequisite for infection with the current strain. If a node fails to become infected with a particular strain, then it cannot become infected with further generations of the disease; therefore, strains that have a low transmissibility significantly reduce the pool of available nodes for future outbreaks. 

To describe the model we index the generations $i\in [1,N]$ as before. Consider a node in the GCC after the first disease has reached its equilibrium, but before the second has emerged. There are two additional types of neighbour that this focal node can be surrounded by: infected (but not by the focal node), and finally, nodes that were directly infected by the focal node in addition to the uninfected substrate nodes. We have to emphasise the particular subset of infected neighbours which the focal node directly infected from those that where infected by one of their other neighbours because the probabilities associated with each state are distinct from one another. 

Now let the second disease emerge, spread and reach its absorbing state, forming its own GCC within the GCC of the first outbreak. Both kinds of infected neighbours from the first process (directly and indirectly infected) give rise to two additional states corresponding to being indirectly or directly infected with strain 2 given direct or indirect infection with strain 1. Therefore, there are now 7 possible neighbour states that might surround the focal vertex in the GCC of the second process (including those already present following strain 1 if further infection with strain 2 did not occur). The third process leads to 15 total neighbour states and each one accounts for the specific details of whom infected whom at each generation. We term a specific combination of direct and indirect infection through the previous generations as an infection history $i_h$. 

The number of new neighbour states at each generation of the branching process is equivalent to the number of leaves of a perfect binary tree, with the total number of neighbour states being the total number of nodes in the tree. This is because each generation branches the current number of maximally coinfected states by a factor of two (accounting for directly and indirectly infected neighbours of the coinfected nodes in generation $i-1$). This means that the total number of neighbour states, $\eta_i$, in the $i$th generation is
\begin{equation}
    \eta_i = 1+\sum_{j=1}^{i}2^j
\end{equation}
comprising all of the states in the previous generation (which did not contract the $i$th strain) in addition to $2^{i-1}$ new states that are directly and indirectly infected, plus the uninfected nodes from the previous generation. Thus, the set of all infection histories for a given generation $i$, $\{h\}_i$, has cardinality $2^{i-1}$; therefore, each generation requires $2^{i-1}$ new $u_{i_h}$ values, with ${i_h}\in\{h\}_i$, $h=1,\dots,2^{i-1}$. A visualisation of this process is provided in appendix \ref{sec:appendixA}.

\subsection{Outbreak size}
\label{sec:b2:obs}

The aim of this section is to define a prescription to obtain the outbreak size of the $i$th epidemic of the coinfection model. Each generation requires $2^{i-1}$ unique $u_{i_h}$ values to be written; each accounting for a particular infection history that a neighbouring node could have. Each $u_{i_h}$ value will then be generated by a self-consistent expression in a similar way to those of the competitive model as
\begin{equation}
   u_{i_h} = \frac{G_1(\mathcal P_{i_h})}{\mathcal Q_{i_h} }\label{eq:u_i_h_type2}
\end{equation}
where $\mathcal P_{i_h}$ is the probability of not obtaining strain $i$ for infection history ${i_h}$ and $\mathcal Q_{i_h}$ is the prior probability that the neighbour has infection history ${i_h}$. Note, $\mathcal P_{{i_h}}$ is analogous to $g_i$ from the competitive model. Thus, it remains to calculate both the prior probabilities and $\mathcal P_{i_h}$. It happens, for a given generation $i$, that the probabilities $\mathcal P_{i_h}$ can be factored; thus, each $\mathcal P_{i_h}$ expression comprises two parts: a multiplying common factor, $\mathcal C_{i}$; and, the unique part of the probability of each specific infection history, $\mathcal H_{i_h}$ such that
\begin{equation}
     \mathcal P_{i_h} = \mathcal C_{i}\mathcal H_{i_h}\label{eq:type2:P}
\end{equation}
Firstly, we calculate $\mathcal C_{i}$, which is simply all of the common terms belonging to each $\mathcal P_{i_h}$, $\forall {i_h}\in\{h\}_i$. Consider each branch point of the collaborative process from the perspective of an infected node as we progress from generation $j-1$ to $j$. Neighbours either do not contract strain-$j$; or they do, from either the focal node or one of their other neighbours. Let $f_j(u_{j_h},v,w)$ be a function that encapsulates the three possible neighbour scenarios and let $u_{j_h}$ be the probability that a neighbour is uninfected by any of its other neighbours by the $j$th strain at this branch point, given that its infection history is $j_h$. We have
\begin{widetext}
\begin{equation}
    f_j(u_{j_h},v,w) = \sum_{l=0}^{k}\binom{k}{l}[u_{j_h}(1-T_j)]^l\sum_{m_{j_h}=0}^{k-l}\binom{k-l}{m_{j_h}}[(1-u_{j_h})v]^{m_{j_h}}[u_{j_h}T_j  w]^{k-l-m_{j_h}}\label{eq:abstract}
\end{equation}
We have indexed the variable $m$ with the infection history for later convenience. Despite the complicated form of this expression it is straightforward to construct each term by considering the probabilities associated with each neighbouring state. In detail: $u_{j_h}(1-T_j)$ is the probability that a neighbour was uninfected by its other neighbours given history $h$ and that the focal node did not transmit strain $j$; $1-u_{j_h}$ is the  probability that a neighbour was already infected and $u_{j_h}T_j$ is the  probability that a neighbour was directly infected by the focal node. The arguments $v$ and $w$ are placeholders that allow the further subdivision of the number of neighbours in a given infected state following the next generation. 

We construct the common factor $\mathcal C_i$ by first constructing a related factor, $\bar{ \mathcal C}_i$, composing this branch-point logic with itself $i$ times and terminating the composition with $v=w=1$ at the deepest levels (i.e. the leaves of the branching process) such that 
\begin{equation}
    \bar{ \mathcal C}_i = f_1(f_2(\dots f_i())))
\end{equation}
The function $\bar{ \mathcal C}_i$ has $2^{i-1}$ arguments. The values of ${j_h}$ in $u_{j_h}$ are given by the particular elements of $\{h\}_j$ $j=1,\dots,i-1$. The common factor for the percolation root (prior to any diseases) is unity, $\bar{\mathcal C}_0=1$; since all strains belong to the same state. Following strain-1 $\bar{\mathcal C}_1$ is given by 
\begin{align}
   \bar{\mathcal C}_1(1,1)=&\ f_1(u_{1_1},1,1)\nonumber\\
   &= u_{1_1}(1-T_1)+1-u_{1_1} + u_{1_1}T_1
\end{align}
for $i=2$ we have 
\begin{align}
     \bar{\mathcal C}_2(1,1,1,1)= &\ f_1(u_{1_1},f_2(u_{2_1},1,1),f_2(u_{2_2},1,1))\nonumber\\ =&\  u_{1_1}(1-T_1)+(1-u_{1_1})(u_{2_1}(1-T_2)+1-u_{2_1} + u_{2_1}T_2) + u_{1_1}T_1(u_{2_2}(1-T_2)+1-u_{2_2} + u_{2_2}T_2)
\end{align}
similarly, for $i=3$ we have
\begin{equation}
     \bar{\mathcal C_3}(\bm 1)=f_1(u_1,f_2(u_{2_1},f_3(u_{3_1},1,1),f_3(u_{3_2},1,1)),f_2(u_{2_2},f_3(u_{3_3},1,1),f_3(u_{3_4},1,1)))
\end{equation}
and so on. An interesting observation is that this expression is always unity and that there are always as many terminating $1$s as there are unique infection histories required for the next generation. We must define another related probability, $\bar f_i$, as
\begin{equation}
    \bar f_j(u_j,v,w) = \sum_{l=0}^k\binom{k}{l}[u_j(1-T_j)]^l\sum_{m=0}^{k-l}\binom{k-l}{m}[(1-u_j)(1-T_j)v]^m[u_jT_j  w]^{k-l-m}\label{eq:abstract2}
\end{equation}
which is equivalent to the $\bar g_i$ expression from section \ref{sec:type1} and is the probability that none of the externally infected neighbours transmitted their infection to the focal node. Given these two functions, we can now build the common terms in the probability that the focal node does not contract the $i$th strain as 
\begin{equation}
    \mathcal C_i = \bar{\mathcal C}_{i-1}(\bm {\bar f_i})
\end{equation}
Thus, $\mathcal C_i $ contains the common terms in the probability that describes the neighbouring states prior to strain $i$, along with the probability that each of those states then fails to transmit strain $i$ itself. For example, the first few values of $\mathcal C_i$ are 
\begin{subequations}
\begin{align}
    \mathcal C_1 =&\  \bar f_1(u_1,1,1)\\
    \mathcal C_2 =&\ \bar{\mathcal C}_1(\bar f(u_{2_1},1,1) ,\bar f(u_{2_2},1,1) )\\
    \mathcal C_3 =&\ \bar{\mathcal C_2}(\bar f_3(u_{3_1},1,1), \bar f_3(u_{3_2},1,1), \bar f_3(u_{3_3},1,1),\bar f_3(u_{3_4},1,1))\label{eq:last_eq_of_3}
\end{align}
\end{subequations}
\begin{figure}[ht!]
  \begin{center}
    \includegraphics[width=1.0\textwidth]{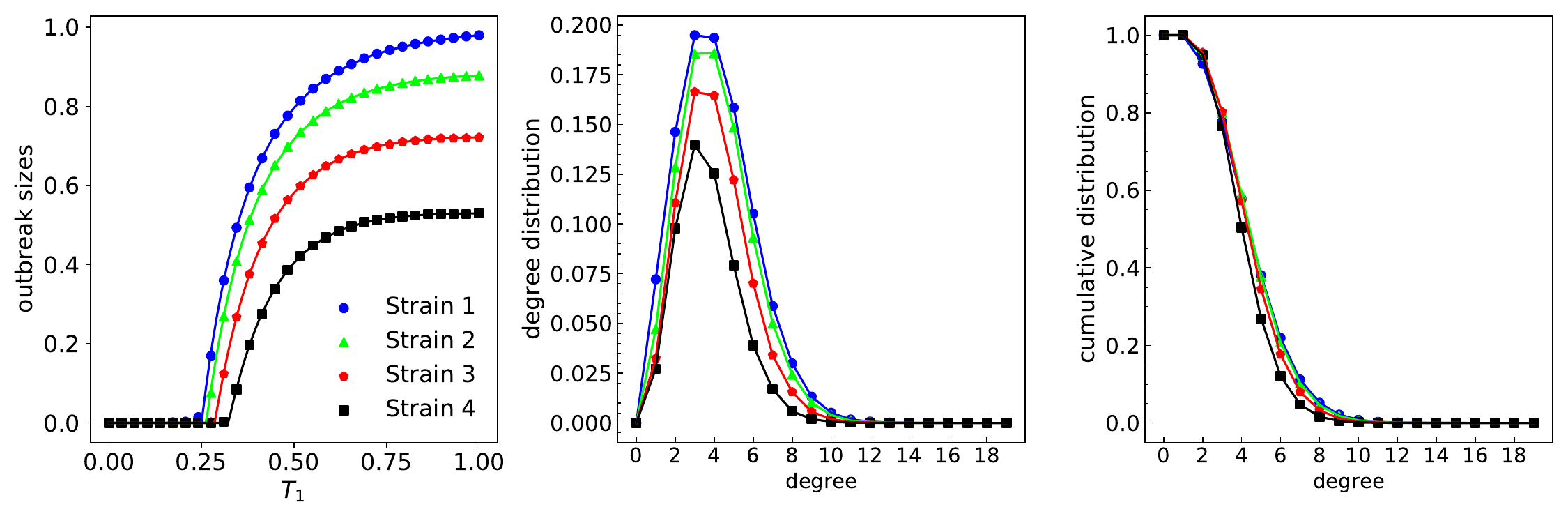}
  \end{center}
  \caption{Four generations of the cooperative branching process with $(T_2,T_3,T_4)=(0.6,0.5,0.45)$ and on an Erd\H{o}s-Renyi network with mean degree $\langle k\rangle=4$. Scatter points are the average of 35 repeats of Monte Carlo simulations over $=30000$ node networks; whilst, solid lines are the theoretical results. Subplot (a) shows the outbreak sizes; (b) shows the degree distribution at $T_1=1$; (c) is the cumulative probability that a node has degree larger than $k$ at $T_1=1$.
  }
  \label{fig:iterated_type2}
\end{figure}
\end{widetext}
With a clear prescription to derive $\mathcal C_i$ for each generation, we must now calculate the probability associated with each infection history $\mathcal H_{i_h}$ in order to finalise the expressions for $\mathcal P_{i_h}$, which in turn we require in order to write self-consistent expressions for each $u_{i_h}$ value in Eq \ref{eq:u_i_h_type2}. To do this, we consider each pathway from the percolation root to the leaves of the tree created by the collaborative branching process. If, at a particular branching point, we progress via direct infection, we require that the focal node was the node that transmitted infection to the neighbour. For this to occur we require that the other neighbours other than the focal node failed to transmit their infection. This occurs with probability $(1-T_j)^{m_r}$, with reference to Eq \ref{eq:abstract}. Similarly, the probability that the neighbour \textit{was} infected by a node other than the focal node is $1-(1-T_j)^{m_r}$. We now see the utility of subscripting $m$ in Eq \ref{eq:abstract} as it allows us to track each particular set of externally infected neighbours over the arguments of $\mathcal C_i$. For instance, there are two infection histories at the start of the second process, strain-1 infected nodes have either been externally infected, ${2_1}$, or directly infected by the focal node, ${2_2}$, such that $\{h\}_2=\{{2_1},{2_2}\}$. The number of externally 1-infected neighbours is given by $m_{1_1}$ and so we have
\begin{align}
    \mathcal H_{2_1} =&\ [1-(1-T_1)^{m_{1_1}}]\\
    \mathcal H_{2_2} =&\ (1-T_1)^{m_{1_1}}
\end{align}
Similarly, a node can obtain strain-3 from one of four different neighbour states: externally-1 and externally-2  infected ($3_1$); externally-1 and directly-2 infected ($3_2$); directly-1 and externally-2 infected ($3_3$), or finally, directly-1 and directly-2 infected ($3_4$). Thus, there are four infection histories to generate with $\{h\}_4= \{3_1,3_2,3_3,3_4\}$. We then write 
\begin{subequations}
 \label{eq:his}
\begin{align}
    \mathcal H_{3_1} = &\  \mathcal H_{2_1}[1-(1-T_2)^{m_{2_1}+m_{2_2}}]\\
    \mathcal H_{3_2} =&\ \mathcal H_{2_1} (1-T_2)^{m_{2_1}+m_{2_2}} \\
    \mathcal H_{3_3} =&\  \mathcal H_{2_2}[1-(1-T_2)^{m_{2_1} + m_{2_2}}]\\
    \mathcal H_{3_4} =&\  \mathcal H_{2_2}(1-T_2)^{m_{2_1}+m_{2_2}}
\end{align}
\end{subequations}
Each history is constructed from the necessary probabilities to create each scenario. For the next generation, each of these unique infection histories are branched into two to give eight potential sources of strain-4.

With these examples, we now have a prescription to write $\mathcal P_{i_h}$ for each potential infection neighbouring state. The last component we require in order to calculate the associated $u_{i_h}$ values in Eq \ref{eq:u_i_h_type2} is the prior probabilities that the neighbour was indeed in that particular state following all of the previous strains, but prior to the $i$th strain itself. It happens that this probability follows a recipe that is simple to compute for each term. We define the following rule: if a neighbour is externally infected at the $j$th strain, we multiply the prior probability by $(1-u_{j_h})$; otherwise, for direct infection, we multiply by $u_{j_h}$ instead. The logic behind this rule is simple: to be directly infected by the focal node, a neighbour must not be infected by their other neighbours; the focal node \textit{must} be the successful infection pathway. Therefore, for the first branch point, the prior probabilities that the neighbour was externally and directly infected, respectively, are given by
\begin{align}
    \mathcal Q_{2_1}=&\ (1-u_{1_1})\\
    \mathcal Q_{2_2}=&\ u_{1_1}
\end{align}
Similarly following the second strain the prior probabilities for the four infection histories are 
\begin{subequations}
\label{eq:priors}
\begin{align}
    \mathcal Q_{3_1} =&\ \mathcal Q_{2_1}(1-u_{2_1})\\
    \mathcal Q_{3_2} =&\ \mathcal Q_{2_1}u_{2_1}\\
    \mathcal Q_{3_3} =&\ \mathcal Q_{2_2}(1-u_{2_2})\\
    \mathcal Q_{3_4} =&\ \mathcal Q_{2_2}u_{2_2}\\
\end{align}
\end{subequations}
With this last component we can now construct the self-consistent expressions required to compute the $u_{i_h}$ values in Eq \ref{eq:u_i_h_type2}. At this point, a useful check is to ensure that the derived components (priors, histories and base term) are correct is to set $\bm {\bar f}_i=\bm 1$. When evaluated at unity, the $u_{i_h}$ values should be equal to one. Both $\mathcal H_{i_h}$ and $\mathcal Q_{i_h}$ can be constructed by multiplying values associated to the nodes of the binary tree of neighbour states from the root to a given leaf state; this is visualised in appendix \ref{sec:appendixA}. 

Next, we require the outbreak size of the $i$th strain, $\mathcal A_i$. It happens that this expression is very simple to construct once we have performed the above work; we simply have to take the expression for $u_{i_h}$ for the maximally indirect, maximally coinfected infection history $i_X$ (i.e. the expression for the history where every node-state was externally infected), remove the prior denominator and replace the $G_1(z)$ generating function with a $G_0(z)$ generating function. We then subtract this value from the previous outbreak size such that 
\begin{equation}
    \mathcal A_i = \mathcal A_{i-1}-G_0(\mathcal P_{i_X})\label{eq:type2main}
\end{equation}
Since $G_0(\mathcal P_{i_X})$ can never be negative, we observe that the outbreak size of each generation can never exceed the size of the previous one. We detail the expressions for the first few generations in appendix \ref{sec:appendixA} and show the percolation results for the first four generations spreading over an Erd\H{o}s-Renyi network in Fig \ref{fig:iterated_type2}. In Fig \ref{fig:iterated_type2} (a) we plot the outbreak sizes of each strain; each exhibits a smaller size and a larger percolation threshold with increasing strain index. In Fig \ref{fig:iterated_type2} (b) the degree distribution of each of the GCC substructures is plotted. The average degree is reduced and the height and variance of each distribution is increasingly reduced and shifted to the left. In Fig \ref{fig:iterated_type2} (c) the cumulative probability that the degree of a node is larger than $k$ is shown for each strain. These results indicate that eventually the spreading of cooperative processes on Erd\H{o}s-Renyi graphs will be limited by the fractured topology of the substrate network available to each strain in addition to the transmissibility of the disease. 

The complete prescription for solving for the outbreak size of the $N$th generation of the cooperative branching process is to hierarchically solve the coupled linear system of equations for each $u_{i_j}$, $i = 1,\dots,N$ given by 
\begin{equation}
u_{i_j} = u_{i_j}(u_{1_1},u_{2_1},\dots,u_{i_{2^{i-1}}};T_1,\dots, T_i)
\end{equation}
for infection histories $j = 1,\dots, 2^{i-1}$, and functional form given by Eq \ref{eq:u_i_h_type2}. More detail on the structure of these expressions in terms of a perfect binary tree is treated in appendix \ref{sec:appendixA} as well as an examination of their solutions.

\subsection{Invasion threshold}

In this section we examine the critical points of the cooperative model, generalising the result of \cite{newman_ferrario_2013} for $N > 2$. As with the competitive percolation, there is an $\mathcal R_{0,i}$ value for each generation or strain. If at any point the outbreak size of a generation is subcritical, then there can be no subsequent outbreaks as coinfection is a strong condition on the proliferation of future strains. However, assuming that the previous $i-1$ strains did indeed cause an $\mathcal O(V)$ outbreak, then there is some point $T_{i,c}$ at which the $i$th strain can also lead to a finite sized propagation if its transmissibility exceeds this value.  

The critical point for the $i$th percolation can be found by applying linear stability analysis on Eq \ref{eq:u_i_h_type2} around the fixed point $\{u^{h}_i\}=1$, which is the trivial root of the system of equations for each generation (see appendix \ref{sec:appendixA}). The critical point of the first strain is identical to the results from section \ref{sec:type1}; however, it is prudent to review this result. Given that $u_{1_1}=1$ at the critical point, we perform a Taylor expansion about $\epsilon_{1_1} = 1-u_{1_1}$ using Eq \ref{eq:u_i_h_type2} and truncate it to 1st order to obtain 
\begin{equation}
    \epsilon_{1_1} \approx 1-G_1(f_1)\big|_{u_{1_1}=1}+G_1'(f_1)f_1'\big|_{u_{1_1}=1}\epsilon_{1_1}
\end{equation}
Rearranging this result, with $G_1(1)=1$ and $f'_1=T_1$, we obtain $ T_{1,c} = {1}/{G_1'(1)}$ in accordance with Eq \ref{eq:criticalTtype1} at $i=1$. For the second strain, we now have two variables to consider depending on the unique infection history of the neighbouring node. The critical point occurs when both $u_{2_1}$ and  $u_{2_2}$ are unity (see appendix \ref{sec:appendixA} for a graphical motivation of this) and we again perform a 1st order Taylor expansion about small parameter $\epsilon_{2_j}=1-u_{2_j}$ to obtain the following coupled system
\begin{align}
    \epsilon_{2_1} \approx\ & \epsilon_{2_1} \frac{\partial F_{2_1}}{\partial u_{2_1}} +\epsilon_{2_2} \frac{\partial F_{2_1}}{\partial u_{2_2}} \nonumber\\
    \epsilon_{2_2} \approx\ & \epsilon_{2_1} \frac{\partial F_{2_2}}{\partial u_{2_1}} +\epsilon_{2_2} \frac{\partial F_{2_2}}{\partial u_{2_2}}
\end{align}
where we have set the functional form of $u_{i_j}$ in Eq \ref{eq:u_i_h_type2} to $u_{i_j}=F_{i_j}$ and evaluate the derivatives at the fixed point $u_{2_1}=u_{2_2}=1$. The derivatives are 
\begin{align}
    \frac{\partial F_{2_1}}{\partial u_{2_1}}&=T_2-G_1'(1-T_1+u_{1_1}T_1)(1-T_1)T_2
    \\
    \frac{\partial F_{2_1}}{\partial u_{2_2}}&=\frac{u_{1_1}T_1}{1-u_{1_1}}\left[1-G_1'(1-T_1+u_{1_1}T_1)\right]T_2
\end{align}
and 
\begin{align}
    \frac{\partial F_{2_2}}{\partial u_{2_1}}&=\frac{G_1'(1-T_1+u_{1_1}T_1)(1-u_{1_1})(1-T_1)}{u_{1_1}}T_2
    \\
    \frac{\partial F_{2_2}}{\partial u_{2_2}}&=G_1'(1-T_1+u_{1_1}T_1)T_1T_2
\end{align}
Thus, we have the following linear system
\begin{equation}
    \bm J\begin{pmatrix}
    u_{2_1}\\
    u_{2_2}
    \end{pmatrix} = \frac{1}{T_{2,c}} \begin{pmatrix}
    u_{2_1}\\
    u_{2_2}
    \end{pmatrix}
\end{equation}
where $\bm J$ is a Jacobian matrix with eigenvalue $1/T_{2,c}$. Following \cite{newman_ferrario_2013} the system has two eigenvalues and by examining the $T_1\rightarrow 1$ limit, the correct eigenvalue is 
\begin{equation}
    T_{2,c} = \frac{2}{\tau+\sqrt{\tau^2-4\Delta}}\label{eq:tracedet}
\end{equation}
where $\tau$ is the trace of $\bm J$ and $\Delta$ is the determinant. 

In the general case we have the following linear system 
\begin{equation}
    \epsilon_{i_j}\approx\sum^{2^{i-1}}_{k=1}\frac{\partial F_{i_j}}{\partial u_{i_k}}\epsilon_{i_k}
\end{equation}
with $i\in [1,n]$ and $j\in [1,2^{i-1}]$. The derivatives are given by 
\begin{equation}
     \frac{\partial F_{i_j}}{\partial u_{i_k}}=\frac{G_1'(\mathcal P_{i_j})}{\mathcal Q (i_j)} \frac{\partial \bar{ \mathcal C}_{i-1}\mathcal H_{i_j}}{\partial \bar f_i}\frac{\partial\bar f_i}{\partial u_{i_k}}\bigg|_{u_{i_k}=1}
\end{equation}
The derivative of the final $\bar f_i$ term is always $\partial \bar f_i=T_i$ meaning that we have a leading factor of $T_i$ multiplying all terms. Thus, we can create the following linear system by simple re-arrangement
\begin{equation}
  \bm J \vec u = \frac{1}{T_{i,c}}\vec u
\end{equation}
where $\vec u =\{u_{1_1},u_{2_1},\dots,u_{i_{2^{i-1}}}\}^{\text{T}}$ and $\bm J $ is a Jacobian matrix with elements $\partial_{ik}F_{i_j}/T_i$ evaluated at the fixed point $\bm u=\{1,1,\dots,1\}$. We then find the eigenvalues by solving $\det(\bm J -\frac{1}{T}\mathbb I)=0$ where $\mathbb I $ is the identity matrix. The characteristic polynomial of an $n\times n$ matrix can be expressed in terms of powers of the trace; however, roots of polynomials of degree five or more are unlikely to yield a closed form solution in general.

\section{Discussion}
\label{sec:discussion}

\subsubsection{$T_i=1$ limit}

Consider the competitive branching process for $N=2$ strains with $T_1=T_2=1$. The first percolation process occupies the entire GCC in the substrate network such that $\mathcal A_1=1-G_0(u_1)$. The outbreak size of the second strain is zero since the fixed point of $\lim_{T_1,T_2\rightarrow 1}u_2=G_1(u_1u_2)/u_1$ is unity and $\mathcal A_2=G_0(u_1) - G_0(u_1u_2)$. This fixed point occurs for all subsequent $u_i$ values for the competitive process and hence all future generations fail to create a finite sized outbreak size.

Consider next the collaborative branching process of $N$ strains with $T_i=1$ for $i=1,\dots,N$. Intuitively, we know that the subsequent generations will occupy all of the edges in the GCC of the first process (the size of which is in turn a function of the topological properties of the substrate network). Thus, we expect
\begin{equation}
    \mathcal A_1 = \mathcal A_i,\qquad \forall i\in [1,N]
\end{equation}
With reference to Eq \ref{eq:type2main} this indicates that $G_0(\mathcal P_{i_X})=0$ where $i_X$ corresponds to the maximally external-coinfected history. This can be seen from the $T_1,T_2\rightarrow 1$ limit of the outbreak size
\begin{align}
    \mathcal A_2=&\ \mathcal A_1-G_0((1-u_{1_1})u_{2_1} \nonumber\\
    &+ u_{1_1}u_{2_2})+G_0(u_{1_1}u_{2_2})
\end{align}
with $  \lim_{T_1,T_2\rightarrow 1} u_{2_1} \rightarrow 0 $ and $ \lim_{T_1,T_2\rightarrow 1} u_{2_2} \rightarrow 1$. For additional generations under the same limit, we observe that all $u_{i_h}$ values have a fixed point at $0$ apart from the history corresponding to the maximally directly coinfected state, which limits to $1$; therefore, this result holds for all generations.

\subsubsection{Scale-free networks}

Let us now consider the application of the competitive and collaborative models to the specific example of disease spreading on scale-free networks. Scale-free networks are good representations of the heavy tailed degree distributions found in many empirical networks \cite{Newman2002SpreadOE,newman_2005_powerlaw}; including human contact networks. Let the degree distribution be given by a power-law of exponent $2< \alpha\leq 3$ with an exponential degree cutoff of $\kappa$ such that 
\begin{equation}
    p_k =
\begin{cases}
0,\qquad &k=0\\
Ck^{-\alpha}e^{-k/\kappa},\qquad &k\geq 1
\end{cases}
\end{equation}
where $C$ is a constant required to normalize the probability distribution to unity. Using the definition of the polylogarithm 
\begin{equation}
    \text{Li}_\alpha(z) = \sum^\infty_{k=0}\frac{z^k}{k^{\alpha}}
\end{equation}
we can write the $G_0(z)$ generating function for this degree distribution as 
\begin{figure}[ht!]
  \begin{center}
    \includegraphics[width=0.40\textwidth]{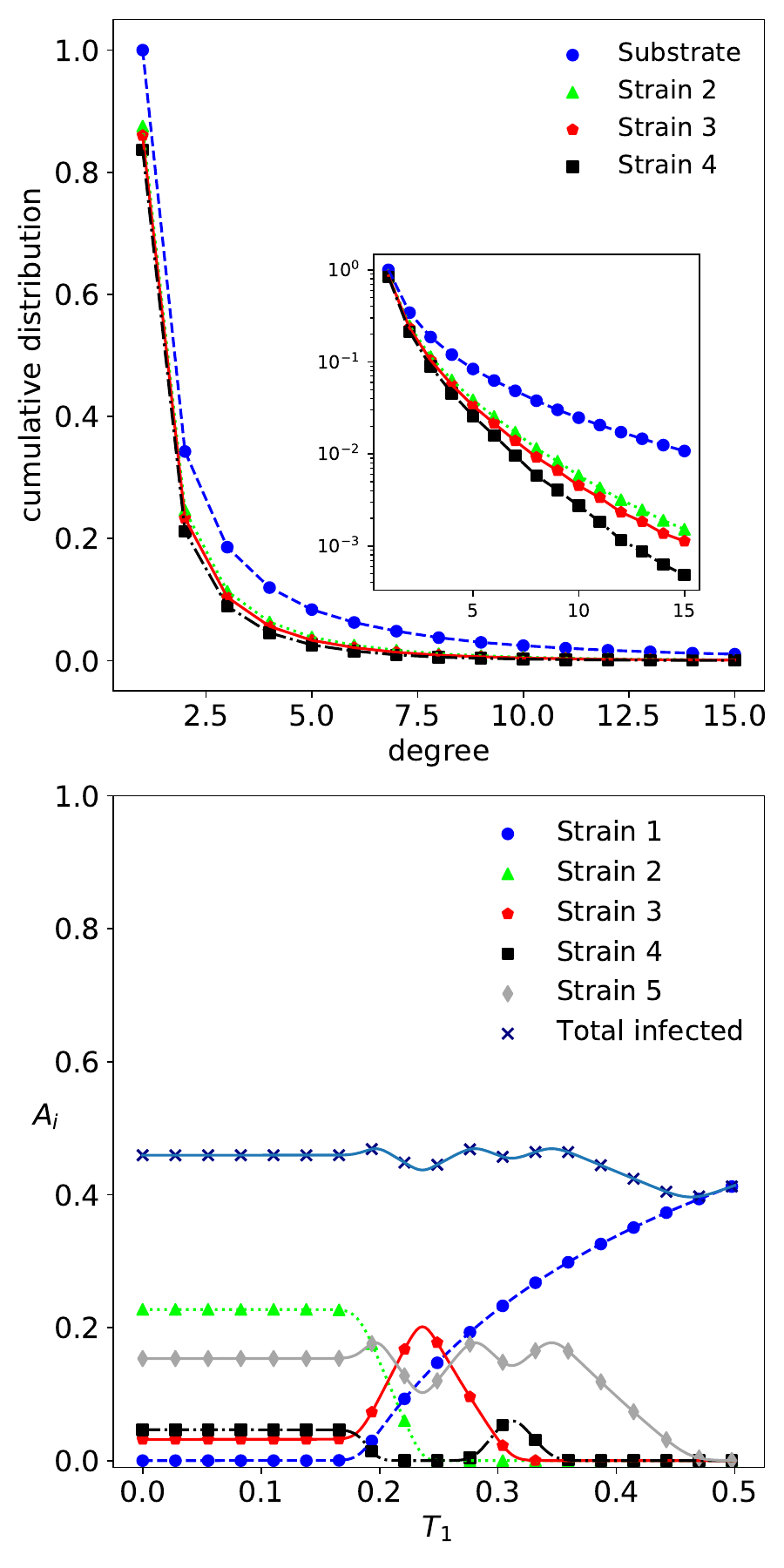}
  \end{center}
  \caption{(top) The cumulative degree distribution of nodes in the substrate network and the GCCs of the RG following competitive bond percolation at $T_1=0.0$. (bottom) The outbreak sizes of the competitive branching process. The parameters are $(T_2,T_3,T_4,T_5)=(0.3,0.5,0.6,1.0)$ on a scale-free network with power-law exponent $\alpha=2$ and $\kappa=20$. Scatter points are the average of 35 repeats of Monte Carlo simulations over $V=30000$ node networks; whilst, solid lines are the theoretical results.
  }
  \label{fig:competitionnscalefree}
\end{figure}
\begin{equation}
    G_0(z) =  \frac{\text{Li}_\alpha(ze^{-1/\kappa})}{ \text{Li}_\alpha(e^{-1/\kappa})}
\end{equation}
whilst for the excess degree distribution we have 
\begin{equation}
    G_1(z) =  \frac{\text{Li}_{\alpha-1}(ze^{-1/\kappa})}{ z\text{Li}_{\alpha-1}(e^{-1/\kappa})}
\end{equation}
Considering the competitive percolation process, the critical point for the $i$th strain is found by applying Eq \ref{eq:criticalTtype1} to this degree distribution; which gives 
\begin{align}
    T_{i,c} =&\  \frac{\text{Li}_{\alpha-1}(e^{-1/\kappa})}{g_{i-1}}\bigg[\text{Li}_{\alpha-2}(g_{i-1} e^{-1/\kappa})\nonumber\\
    &- \frac{1}{g_{i-1}}\text{Li}_{\alpha-1}(g_{i-1}e^{-1/\kappa})\bigg]^{-1}
\end{align}
Thus, with $g_0=1$, the critical point of the first strain of the competitive process reduces to the expression found by Newman \cite{Newman2002SpreadOE} \begin{equation}
    T_{1,c} = \frac{\text{Li}_{\alpha-1}(e^{-1/\kappa})}{\text{Li}_{\alpha-2}(e^{-1/\kappa})-\text{Li}_{\alpha-1}(e^{-1/\kappa})}
\end{equation}
In the high-degree limit with $\kappa\rightarrow\infty$, the epidemic threshold of strain 1 occurs at $T_{1,c}=0$; since, the second moment of the degree distribution diverges \cite{PhysRevLett.86.3200}. However, the critical point of strain 2 is non-zero and is given by 
\begin{equation}
    \lim_{\kappa
    \rightarrow \infty}T_{2,c}\approx\frac{1}{\zeta(\alpha)}\frac{\text{Li}_{\alpha-1}(g_1)}{g_1}\label{eq:T2comp}
\end{equation}
where $\zeta(\alpha)$ is the Riemann $\zeta$-function. This indicates the rapid fracture of the RG by the first generation and the subsequent loss of the scale-free property \cite{newman_2005}. This is because the GCC of the first competitive process targets the high-degree sites of the network, removing the power-law distribution of degrees. In the limit that $T_1\rightarrow 1$, Eq \ref{eq:T2comp} diverges, exceeding the coexistence threshold and strain 2 fails to spread on the RG. In Fig \ref{fig:competitionnscalefree} (bottom) we show the cumulative degree distribution (top) and the outbreak sizes (bottom) of the first five strains of the competitive process on a scale-free network with power-law exponent $\alpha=2$ and degree cutoff $\kappa=20$ for transmissibilities $(T_2,T_3,T_4,T_5)=(0.3,0.5,0.6,1.0)$. In this experiment, the total infected fraction of the network is fairly constant $\approx 0.5$; however, the outbreak sizes of each strain are strong, multimodal functions of the available space left to spread in the RG of the previous strains. These disease dynamics show that each generation of a disease can generate a finite GCC of its own in the presence of other outbreaks. Fig \ref{fig:competitionnscalefree} (top - inset) shows the same data as the main panel on a semi-log plot. Each generation leads to an increased fracturing of the network and a transition from scale-free behaviour (linearity on a log-log plot) to the emergence of an exponential relationship (linearity on a semi-log plot) with increasing strain index.

\begin{widetext}
Turning now to the cooperative process, the general expression for the critical transmissibility of the strain 2 was given in  \cite{newman_ferrario_2013} and is shown in Eq \ref{eq:tracedet}. For the scale-free degree distribution we have the trace and determinant given by 
\begin{align}
    \tau=&\ 1+(2T_1-1) \frac{\text{Li}_{\alpha-1}(e^{-1/\kappa})}{\bar {f}_1 }\bigg[\text{Li}_{\alpha-2}(\bar {f}_1  e^{-1/\kappa})\nonumber- \frac{1}{\bar {f}_1 }\text{Li}_{\alpha-1}(\bar {f}_1 e^{-1/\kappa})\bigg]^{-1}
\end{align}
and 
\begin{align}
    \Delta =&\ T_1^2\frac{\text{Li}_{\alpha-1}(e^{-1/\kappa})}{\bar {f}_1 }\bigg[\text{Li}_{\alpha-2}(\bar {f}_1  e^{-1/\kappa})\nonumber- \frac{1}{\bar {f}_1 }\text{Li}_{\alpha-1}(\bar {f}_1 e^{-1/\kappa})\bigg]^{-1}
\end{align}
respectively. At $\kappa\rightarrow\infty$, both $\tau\rightarrow \infty$ and $\Delta\rightarrow\infty$ and so $T_{2_c}=0$ from Eq \ref{eq:tracedet}. Since the critical point of subsequent strains will always contain this factor, their critical points also vanish in this limit. Thus, cooperative diseases on scale-free networks will theoretically always spread; regardless of measures that lower the strain transmissibility. In Fig \ref{fig:coinfectionscalefree} we show the percolation properties of the first four strains of the cooperative branching process on a scale-free network with power-law exponent $\alpha=2.0$ and degree cutoff $\kappa=20$; the transmission probability of each strain is $(T_2,T_3,T_4)=(0.6,0.5,0.45)$. Fig \ref{fig:coinfectionscalefree} (a) shows the outbreak fractions become successively smaller with increasing generations. In Fig \ref{fig:coinfectionscalefree} (b) we see the degree distributions of GCCs formed by each generation of the process at $T_1=1.0$. The inset shows (from a repeated experiment with $\kappa=200$) that the heavy tail of the power-law is preserved following each percolation, whilst the probability of choosing a low degree node becomes increasingly smaller. In Fig \ref{fig:coinfectionscalefree} (c) the cumulative probability that a node has degree larger than $k$ within the GCC structures shows the removal of the low degree nodes with each generation of the percolation. This is in direct contrast to the fracture process that occurs for Erd\H{o}s-Renyi networks (see Fig \ref{fig:iterated_type2} (c)), where the cumulative probability \textit{decreased} with generation index under cooperative percolation. The preservation of the higher-degree core in the scale-free network indicates the self-similarity of the GCC substructures under cooperative percolation. 

\begin{figure}[ht!]
  \begin{center}
    \includegraphics[width=1.0\textwidth]{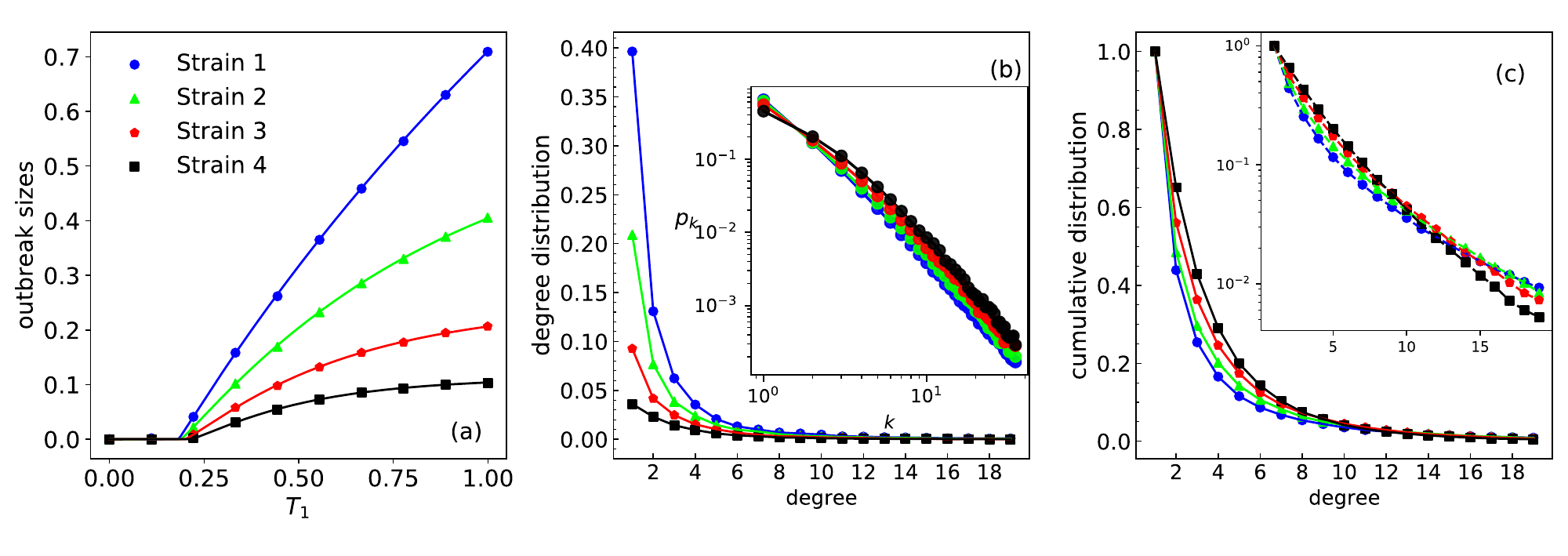}
  \end{center}
  \caption{Four generations of the cooperative branching process with $(T_2,T_3,T_4)=(0.6,0.5,0.45)$ on a scale-free network with power-law exponent $\alpha=2$ and $\kappa=20$. Scatter points are the average of 35 repeats of Monte Carlo simulations over $V=500000$ node networks; whilst, solid lines are the theoretical results. Subplot (a) shows the outbreak sizes, (b) shows the degree distribution at $T_1=1$, the inset is a log-log plot with $\kappa=200$; whilst, (c) is the cumulative probability that a node has degree larger than $k$ in each GCC at $T_1=1$; the inset shows the same data on a logarithmic scale for $\kappa=20$.
  }
  \label{fig:coinfectionscalefree}
\end{figure}
\end{widetext}

\section{Conclusion}
\label{sec:conclusion}

In this paper we have introduced two exact models of generational bond percolation on complex networks. The outbreak sizes as well as the critical points of the models were solved for and topological properties of the resulting graph structures were examined.

Competitive branching processes are defined as a perfect cross-immunity model; each generation spreading on the RG created by the previous generations. It was shown, under the constraint that each disease adopts its critical transmissibility, that there is an evolutionary pressure on each generation to become more transmissible; however, the expected outbreak size becomes increasingly smaller. Collaborative branching processes are defined as a complete coinfection model; each generation spreading on the GCC created by all of the previous strains. It was shown that the outbreak size of each strain is bound by the size of the preceding epidemic.

Both of the stochastic branching processes studied in this paper increasingly fracture the substrate network as they unfold. This eventually leads to the loss of global connectivity among the nodes of the network and the eventual burn-out of each process. We examined the resulting graph structures from both competitive and cooperative percolation types for two common degree distributions: Erd\H{o}s-Renyi and power-law networks. The RG structures created by the competitive process led to non-trivial outbreak sizes that can exhibit multiple turning points depending on the transmissibility of previous strains. Additionally, for both Erd\H{o}s-Renyi and scale-free networks the total outbreak size was found to be a non-monotonic function of the proceeding fracture process. 

We found that the GCC structures created through cooperative percolation on Erd\H{o}s-Renyi networks fracture through a gradual loss of high degree sites. Scale-free networks, in contrast, produce self-similar outbreaks supported by a power-law core. 

Whilst these scenarios are illuminating in their own right, this work is an important step towards an $N$-strain partial immunity model, whereby a strain could infect \textit{all} nodes irregardless of their infection history. Recent work by Mann \textit{et al.} \cite{PhysRevE.104.024303} has shown that bond percolation can also be mapped to a partial immunity model for $N=2$. In this case, the strict criteria concerning the disease couplings required for the \textit{complete} cross-immunity and \textit{complete} coinfection models are relaxed. Importantly, such a model would not burn out. This work can be generalised in many different ways such as: networks with clustering \cite{PhysRevE.104.024304}, including hypergraphs as well as modular systems or multitype networks \cite{oz_rubinstein_safra_2022,PhysRevE.103.012309}. These processes could also prove useful to the understanding of other dynamical processes on networks, such as synchronisation. Finally, a strong condition on the success of the model is the temporal separation of each generation; we have not addressed any dynamical features of the spreading processes, such as their time-scales or growth rates, which we leave for future work.

\section{ACKNOWLEDGMENTS}

This work was partially supported by the UK Engineering and Physical Sciences Research Council under grant number EP/N007565/1 (Science of Sensor Systems Software).

\subsection*{References}
\bibliography{bib}

\appendix 

\begin{widetext}
\section{Outbreak sizes for collaborative branching processes}
\label{sec:appendixA}

In this appendix we describe how to use the prescription to obtain the outbreak sizes of the first few generations of the collaborative branching process.  

\subsection*{Strain 1}

The prescription for the outbreak size of strain 1 is as follows. First we calculate $\mathcal C_1 =\bar f_1(u_1,1,1)$. A node in the GCC has three neighbour states: uninfected, infected (externally by other nodes) and infected (directly by the focal node); of which, only the two infected states play an active role in the spread of further generations. The probability that an edge between the focal node and an externally infected neighbour failed to transmit strain 1 is $1-T_1$. Thus, the probability that all $m_{1_1}$ externally infected neighbours failed is $(1-T_1)^{m_{1_1}}$. The probability that the neighbour was uninfected is then 
\begin{equation}
    u_{1_1} = \sum_{k=0}^\infty q_k[u_{1_1}(1-T_1) + (1-u_{1_1})(1-T_1) + u_{1_1}T_1]^{k}
\end{equation}
where $q_k=(k+1)p_{k+1}/\langle k \rangle$ from Eq \ref{eq:g1g1}. With this probability, we now write the outbreak size as
\begin{equation}
    \mathcal A_1 = 1 - \sum_{k=0}^\infty p_k[u_{1_1}(1-T_1) + (1-u_{1_1})(1-T_1) + u_{1_1}T_1]^k 
\end{equation}
The bracket reduces to Eq \ref{eq:gbar}; thus, both competitive and collaborative formulations are in agreement for the first percolation process. The solution $u_{1_1}$ can also be visualised graphically, thus confirming the generating function is convex on the unit interval and that Jensen's inequality can be applied, Fig \ref{fig:non-trivialroot}.

\begin{figure}[ht!]
  \begin{center}
    \includegraphics[width=0.35\textwidth]{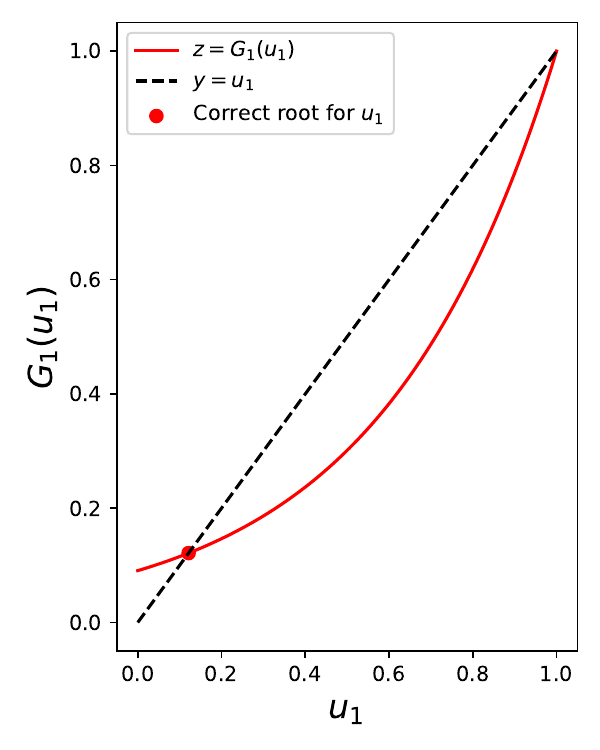}
  \end{center}
  \caption{The graphical solution \cite{newman_2019} of the generating function for strain 1 at $T_1=0.6$. Plotted are $y=u_{1_1}$ against $u_{1_1}$ as well as the value of the generating function $z_1=G_1(1-T_1+u_{1_1}T_1)$. The intersection of $y=u_{1_1}$ and $z_1$ corresponds to the root. We notice that the trivial root, $u_{1_1}$ is also a solution. The scatter point is the result of fixed point iteration.
  }
  \label{fig:non-trivialroot}
\end{figure}

\subsection*{Strain 2}

For strain 2 there are two possible infection histories that a neighbour might have: either the focal node infected it directly or it was externally infected. The probability that the focal node doesn't get strain 2 from each of these neighbour states is $u_{2_1}$ and $u_{2_2}$, respectively. The common factor is given by  
\begin{equation}
    \mathcal C_2 = \bar{\mathcal C}_1(\bar f(u_{2_1},1,1) ,\bar f(u_{2_2},1,1))
\end{equation}
which is simply 
\begin{equation}
   \mathcal C_2 = \sum_{l=0}^k\binom{k}{l}[u_{1_1}(1-T_1)]^l\sum_{m_{1_1}=0}^{k-l}\binom{k-l}{m_{1_1}}[(1-u_{1_1})\bar f(u_{2_1})]^{m_{1_1}}[u_{1_1}T_1  \bar f(u_{2_2})]^{k-l-m_{1_1}}
\end{equation}
where we have dropped the $1$s in the function arguments of the $\bar f$ functions. Following the prescription, the history of $u_{2_1}$ is $\mathcal H_{2_1}=[1-(1-T_1)^{m_{1_1}}]$ whilst the history of $u_{2_2} $ is $\mathcal H_{2_2}=(1-T_1)^{m_{1_1}}$. Since the $u_{2_1}$ history branches from the $1-u_{1_1}$ compartment, the prior probability is simply $ \mathcal Q_{2_1}=1-u_{1_1}$; whilst $  \mathcal Q_{2_2}=u_{1_1}$. Thus, we have 
\begin{align}
u_{2_1} = &\ \frac{1}{ \mathcal Q_{2_1}} \sum^\infty_{k=0}q_k \mathcal C_2 \mathcal H_{2_1}\\
u_{2_2} = &\ \frac{1}{\mathcal Q_{2_2}} \sum^\infty_{k=0}q_k \mathcal C_2 \mathcal H_{2_2}
\end{align}
As we insert the infection histories into this expression we observe that externally infected histories lead to multiple brackets as the generating function acts on each term. For instance
\begin{align}
    u_{2_1} = &\ \frac{1}{\mathcal Q_{2_1}} \sum^\infty_{k=0}q_k \mathcal C_2 [1-(1-T_1)^{m_{1_1}}]\\
    = &\frac{1}{\mathcal Q_{2_1}} \sum^\infty_{k=0}q_k \mathcal C_2 -\frac{1}{\mathcal Q_{2_1}} \sum^\infty_{k=0}q_k \mathcal C_2(1-T_1)^{m_{1_1}}\\
    =&\ \frac{G_1(C_2)}{\mathcal Q_{2_1}}- \frac{G_1(C_2(1-T_1)^{m_{1_1}})}{\mathcal Q_{2_1}}\\
    =&\ \frac{G_1(u_{1_1}(1-T_1) + (1-u_{1_1})\bar f_2(u_{2_1})+u_{1_1}T_1\bar f_2(u_{2_2}))}{1-u_{1_1}} - \frac{G_1(u_{1_1}(1-T_1) + (1-u_{1_1})(1-T_1)\bar f_2(u_{2_1})+u_{1_1}T_1\bar f_2(u_{2_2}))}{1-u_{1_1}} 
\end{align}
This doesn't occur for $u_{2_2}$ since the infection history folds into $\mathcal C_2$. 
\begin{equation}
    u_{2_2} = \frac{G_1(u_{1_1}(1-T_1) + (1-u_{1_1})(1-T_1)\bar f_2(u_{2_1})+u_{1_1}T_1\bar f_2(u_{2_2}))}{u_{1_1}}
\end{equation}
We exhibit the graphical solution for these coupled equations in Fig \ref{fig:non-trivialroot2}.
\begin{figure}[ht!]
  \begin{center}
    \includegraphics[width=0.99\textwidth]{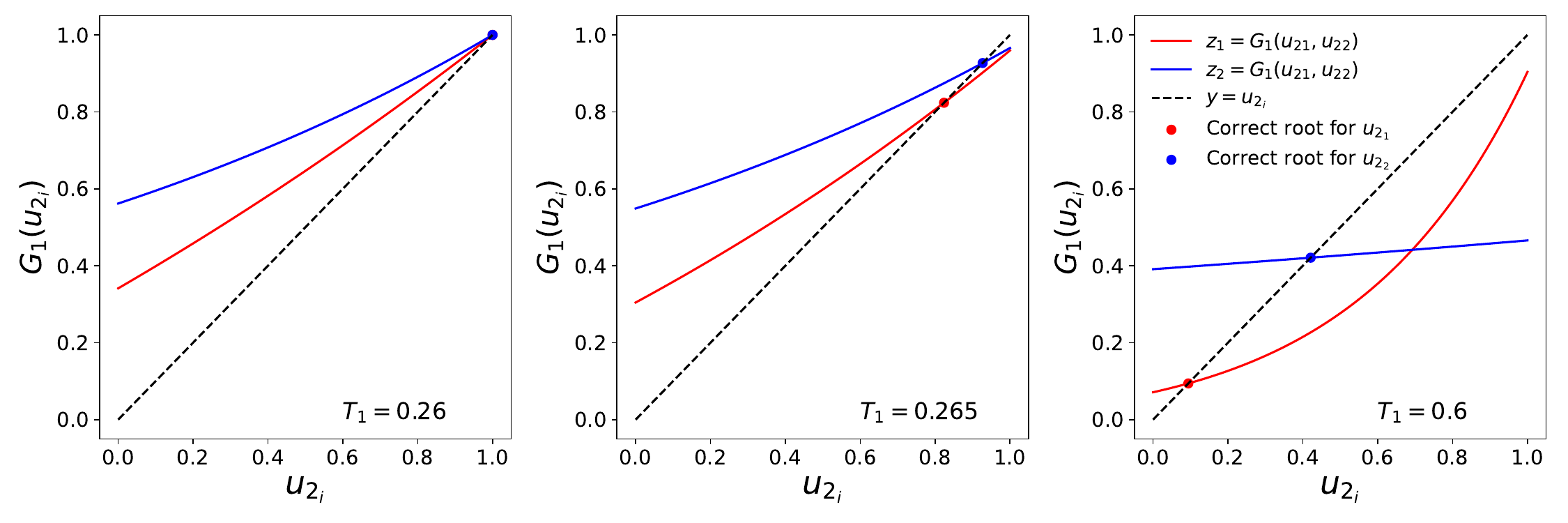}
  \end{center}
  \caption{The graphical solution of the generating functions for strain 2 at three different $T_1$ values with $T_2=0.6$. Plotted are $y=u_{2_i}$ for $i=1,2$ against $u_{2_i}$ as well as the value of the generating functions $z_1=G_1(\mathcal C_2\mathcal H_{2_1})/(1-u_{1_1})$ and $z_2=G_1(\mathcal C_2\mathcal H_{2_2})/u_{1_1}$. Each $z_i$ varies $u_{2_i}$ whilst the other value is held fixed at the correct root. The intersection of $y=u_{2_i}$ and $z_i$ corresponds to the root, which is also marked with a scatter point. We notice that the trivial root of $u_{2_1}=u_{2_2}=1$ is no longer shown as the system moves away from the critical point when the GCC first forms. This is also graphical motivation for finding the critical point from a Taylor series around the trivial root. We also notice the increase (loss) of convexity in $u_{2_1}$ ($u_{2_2}$) as we increase $T_1$. This indicates the increasing (decreasing) importance of the $u_{2_1}$ ($u_{2_2}$) branch to the formation of the GCC at larger transmissibilities.  
  }
  \label{fig:non-trivialroot2}
\end{figure}

\subsection*{Strain 3}
The first step is to write the common factor $\bar{\mathcal C}_2(\bm 1)$, (see Eq \ref{eq:last_eq_of_3}); this is given by 
\begin{align}
    \bar{\mathcal C}_2 =&\  \sum_{l=0}^k\binom{k}{l}[u_{1_1}(1-T_1)]^l\nonumber\\
    &\ \times\sum_{m_{1_1}=0}^{k-l}\binom{k-l}{m_{1_1}}[1-u_{1_1}]^{m_{1_1}}\sum_{a=0}^{m_{1_1}}\binom{m_{1_1}}{a}[u_{2_1}(1-T_2)]^a\sum_{m_{2_1}=0}^{m_{1_1}-a}\binom{m_{1_1}-a}{m_{2_1}}[1-u_{2_1}]^{m_{2_1}}[u_{2_1}T_2]^{m_{1_1}-a-m_{2_1}}\nonumber\\
    &\ \times [u_{1_1}T_1]^{k-l-m_{1_1}}\sum_{c=0}^{k-l-m_{1_1}}\binom {k-l-m_{1_1}}{c}[u_{2_2}(1-T_2)]^c\sum_{m_{2_2}=0}^{k-l-m_{1_1}-c}\binom{k-l-m_{1_1}-c}{m_{2_2}}[1-u_{2_2}]^{m_{2_2}}[u_{2_2}T_2]^{k-l-m_{1_1}-c-m_{2_2}}
\end{align}
We then form $\mathcal C_3 = \bar{\mathcal C}_2 (\bar f_3(u_{3_1},1,1), \bar f_3(u_{3_2},1,1), \bar f_3(u_{3_3},1,1),\bar f_3(u_{3_4},1,1))$ as
\begin{align}
    \mathcal C_3 =&\  \sum_{l=0}^k\binom{k}{l}[u_{1_1}(1-T_1)]^l\nonumber\\
    &\ \times\sum_{m_{1_1}=0}^{k-l}\binom{k-l}{m_{1_1}}[1-u_{1_1}]^{m_{1_1}}\sum_{a=0}^{m_{1_1}}\binom{m_{1_1}}{a}[u_{2_1}(1-T_2)]^a\sum_{m_{2_1}=0}^{m_{1_1}-a}\binom{m_{1_1}-a}{m_{2_1}}[(1-u_{2_1})\bar f_3(u_{3_1})]^{m_{2_1}}[u_{2_1}T_2\bar f_3(u_{3_2})]^{m_{1_1}-a-m_{2_1}}\nonumber\\
    &\ \times [u_{1_1}T_1]^{k-l-m_{1_1}}\sum_{c=0}^{k-l-m_{1_1}}\binom {k-l-m_{1_1}}{c}[u_{2_2}(1-T_2)]^c\sum_{m_{2_2}=0}^{k-l-m_{1_1}-c}\binom{k-l-m_{1_1}-c}{m_{2_2}}[(1-u_{2_2})\bar f_3(u_{3_3})]^{m_{2_2}}[u_{2_2}T_2\bar f_3(u_{3_4})]^{s}
\end{align}
where we have dropped the 1s from the $\bar f$ arguments and simplified $s=k-l-m_{1_1}-c-m_{2_2}$ for brevity. We define the set $\{u_{3_j}\}$ $j=1,\dots,4$, as the set of probabilities that each neighbour fails to transmit strain-3 to the focal node given their particular history. The probabilities, $\mathcal H_{3_j}$, of each of these unique histories occurring were defined in Eqs \ref{eq:his}. Using Eq \ref{eq:type2:P} we now furnish $\mathcal C_3$ with these probabilities to obtain
\begin{subequations}
\begin{align}
    \mathcal P_{3_1} =&\  \sum_{l=0}^k\binom{k}{l}[u_{1_1}(1-T_1)]^l\sum_{m_{1_1}=0}^{k-l}\binom{k-l}{m_{1_1}}[1-u_{1_1}]^{m_{1_1}}\sum_{a=0}^{m_{1_1}}\binom{m_{1_1}}{a}[u_{2_1}(1-T_2)]^a\sum_{m_{2_1}=0}^{m_{1_1}-a}\binom{m_{1_1}-a}{m_{2_1}}[(1-u_{2_1})\bar f_3(u_{3_1})]^{m_{2_1}}\nonumber\\
    &\ \times[u_{2_1}T_2\bar f_3(u_{3_2})]^{m_{1_1}-a-m_{2_1}} [u_{1_1}T_1]^{k-l-m_{1_1}}\sum_{c=0}^{k-l-m_{1_1}}\binom {k-l-m_{1_1}}{c}[u_{2_2}(1-T_2)]^c\nonumber\\
    &\ \times\sum_{m_{2_2}=0}^{k-l-m_{1_1}-c}\binom{k-l-m_{1_1}-c}{m_{2_2}}[(1-u_{2_2})\bar f_3(u_{3_3})]^{m_{2_2}}[u_{2_2}T_2\bar f_3(u_{3_4})]^{s}[1-(1-T_1)^{m_{1_1}}][1-(1-T_2)^{m_{2_1}+m_{2_2}}]
\\ \nonumber \\
    \mathcal P_{3_2} =&\  \sum_{l=0}^k\binom{k}{l}[u_{1_1}(1-T_1)]^l\sum_{m_{1_1}=0}^{k-l}\binom{k-l}{m_{1_1}}[1-u_{1_1}]^{m_{1_1}}\sum_{a=0}^{m_{1_1}}\binom{m_{1_1}}{a}[u_{2_1}(1-T_2)]^a\nonumber\\
    &\ \times\sum_{m_{2_1}=0}^{m_{1_1}-a}\binom{m_{1_1}-a}{m_{2_1}}[(1-u_{2_1})(1-T_2)\bar f_3(u_{3_1})]^{m_{2_1}}[u_{2_1}T_2\bar f_3(u_{3_2})]^{m_{1_1}-a-m_{2_1}} [u_{1_1}T_1]^{k-l-m_{1_1}}\nonumber\\
    &\ \times\sum_{c=0}^{k-l-m_{1_1}}\binom {k-l-m_{1_1}}{c}[u_{2_2}(1-T_2)]^c\sum_{m_{2_2}=0}^{k-l-m_{1_1}-c}\binom{k-l-m_{1_1}-c}{m_{2_2}}[(1-u_{2_2})(1-T_2)\bar f_3(u_{3_3})]^{m_{2_2}}\nonumber\\
    &\ \times[u_{2_2}T_2\bar f_3(u_{3_4})]^{s}[1-(1-T_1)^{m_{1_1}}]
\\ \nonumber \\
    \mathcal P_{3_3} =&\  \sum_{l=0}^k\binom{k}{l}[u_{1_1}(1-T_1)]^l\sum_{m_{1_1}=0}^{k-l}\binom{k-l}{m_{1_1}}[(1-u_{1_1})(1-T_1)]^{m_{1_1}}\sum_{a=0}^{m_{1_1}}\binom{m_{1_1}}{a}[u_{2_1}(1-T_2)]^a\nonumber\\
    &\ \times\sum_{m_{2_1}=0}^{m_{1_1}-a}\binom{m_{1_1}-a}{m_{2_1}}[(1-u_{2_1})\bar f_3(u_{3_1})]^{m_{2_1}}[u_{2_1}T_2\bar f_3(u_{3_2})]^{m_{1_1}-a-m_{2_1}} [u_{1_1}T_1]^{k-l-m_{1_1}}\nonumber\\
    &\ \times\sum_{c=0}^{k-l-m_{1_1}}\binom {k-l-m_{1_1}}{c}[u_{2_2}(1-T_2)]^c\sum_{m_{2_2}=0}^{k-l-m_{1_1}-c}\binom{k-l-m_{1_1}-c}{m_{2_2}}[(1-u_{2_2})\bar f_3(u_{3_3})]^{m_{2_2}}\nonumber\\
    &\ \times[u_{2_2}T_2\bar f_3(u_{3_4})]^{s} [1-(1-T_2)^{m_{21}+m_{22}}]
\\ \nonumber \\
    \mathcal P_{3_4} =&\  \sum_{l=0}^k\binom{k}{l}[u_{1_1}(1-T_1)]^l\sum_{m_{1_1}=0}^{k-l}\binom{k-l}{m_{1_1}}[(1-u_{1_1})(1-T_1)]^{m_{1_1}}\sum_{a=0}^{m_{1_1}}\binom{m_{1_1}}{a}[u_{2_1}(1-T_2)]^a\nonumber\\
    &\ \times\sum_{m_{2_1}=0}^{m_{1_1}-a}\binom{m_{1_1}-a}{m_{2_1}}[(1-u_{2_1})(1-T_2)\bar f_3(u_{3_1})]^{m_{2_1}}[u_{2_1}T_2\bar f_3(u_{3_2})]^{m_{1_1}-a-m_{2_1}}\nonumber\\
    &\ \times [u_{1_1}T_1]^{k-l-m_{1_1}}\sum_{c=0}^{k-l-m_{1_1}}\binom {k-l-m_{1_1}}{c}[u_{2_2}(1-T_2)]^c\sum_{m_{2_2}=0}^{k-l-m_{1_1}-c}\binom{k-l-m_{1_1}-c}{m_{2_2}}[(1-u_{2_2})(1-T_2)\bar f_3(u_{3_3})]^{m_{2_2}}\nonumber\\
    &\ \times[u_{2_2}T_2\bar f_3(u_{3_4})]^{s}
\end{align}
\end{subequations}
Next, we construct the prior probabilities that the neighbour in question was in the desired state; these are shown in Eqs \ref{eq:priors} for each history. Lastly, we use Eq \ref{eq:u_i_h_type2} to finalise the self-consistent expressions for the $u_{i_h}$ values using
\begin{equation}
    u_{i_h} = \frac{1}{\mathcal Q_{i_{h}}}\sum^\infty_{k=0}q_k\mathcal P_{i_h}
\end{equation}
In some cases, this expression is straightforward to compute; however, those histories obtained from external coinfection lead to multiple $G_1(z)$ terms when the brackets are multiplied out. For instance $\mathcal H_{3_1}$ behaves as follows
\begin{equation}
    [1-(1-T_1)^{m_{1_1}}][1-(1-T_2)^{m_{2_1}+m_{2_2}}] = 1 - (1-T_1)^{m_{1_1}}-(1-T_2)^{m_{2_1}+m_{2_2}} +(1-T_1)^{m_{1_1}}(1-T_2)^{m_{2_1}+m_{2_2}}
\end{equation}
When $\mathcal C_2$ acts on this expression we generate 4 brackets. We find 
\begin{subequations}
\begin{alignat}{2}
    u_{3_1} =&\frac{1}{(1-u_{1_1})(1-u_{2_1})}G_1(u_{1_1}(1-T_1)+ (1-u_{1_1})[u_{2_1}(1-T_2) + (1-u_{2_1})\bar f_3(u_{3_1}) + u_{2_1}T_2\bar f_3(u_{3_2})]\nonumber\\
    &+ u_{1_1}T_1[u_{2_2}(1-T_2) + (1-u_{2_2})\bar f_3(u_{3_3}) + u_{2_2}T_2\bar f_3(u_{3_4})]) \nonumber\\
    &- 
    \frac{1}{(1-u_{1_1})(1-u_{2_1})}G_1(u_{1_1}(1-T_1)+ (1-u_{1_1})[u_{2_1}(1-T_2) + (1-u_{2_1})(1-T_2)\bar f_3(u_{3_1}) + u_{2_1}T_2\bar f_3(u_{3_2})]\nonumber\\
    &+ u_{1_1}T_1[u_{2_2}(1-T_2) + (1-u_{2_2})(1-T_2)\bar f_3(u_{3_3}) + u_{2_2}T_2\bar f_3(u_{3_4})]) \nonumber\\
    &- 
    \frac{1}{(1-u_{1_1})(1-u_{2_1})}G_1(u_{1_1}(1-T_1)+ (1-u_{1_1})(1-T_1)[u_{2_1}(1-T_2) + (1-u_{2_1})\bar f_3(u_{3_1}) + u_{2_1}T_2\bar f_3(u_{3_2})]\nonumber\\
    &+ u_{1_1}T_1[u_{2_2}(1-T_2) + (1-u_{2_2})\bar f_3(u_{3_3}) + u_{2_2}T_2\bar f_3(u_{3_4})]) 
    \nonumber\\
    &+ 
    \frac{1}{(1-u_{1_1})(1-u_{2_1})}G_1(u_{1_1}(1-T_1)+ (1-u_{1_1})(1-T_1)[u_{2_1}(1-T_2) + (1-u_{2_1})(1-T_2)\bar f_3(u_{3_1}) + u_{2_1}T_2\bar f_3(u_{3_2})]\nonumber\\
    &+ u_{1_1}T_1[u_{2_2}(1-T_2) + (1-u_{2_2})(1-T_2)\bar f_3(u_{3_3}) + u_{2_2}T_2\bar f_3(u_{3_4})]) 
\\ \nonumber \\
    u_{3_2} =& \frac{1}{(1-u_{1_1})u_{2_1}}G_1(
u_{1_1}(1-T_1) 
+ (1-u_{1_1})[u_{2_1}(1-T_2) + (1-u_{2_1})(1-T_2)\bar f_3(u_{3_1}) + u_{2_1}T_2\bar f_3(u_{3_2})]\nonumber\\
&
+ u_{1_1}T_1[u_{2_2}(1-T_2) + (1-u_{2_2})(1-T_2)\bar f_3(u_{3_3}) + u_{2_2}T_2\bar f_3(u_{3_4})]
    ) 
    -
    \frac{1}{(1-u_{1_1})u_{2_1}}G_1(
    u_{1_1}(1-T_1) \nonumber\\
&+ (1-u_{1_1})(1-T_1)[u_{2_1}(1-T_2) + (1-u_{2_1})(1-T_2)\bar f_3(u_{3_1}) + u_{2_1}T_2\bar f_3(u_{3_2})]
+ u_{1_1}T_1[u_{2_2}(1-T_2) \nonumber\\
&+ (1-u_{2_2})(1-T_2)\bar f_3(u_{3_3}) + u_{2_2}T_2\bar f_3(u_{3_4})]
    ) 
    \\ \nonumber \\
    u_{3_3} =& \frac{1}{u_{1_1}(1-u_{2_2})} G_1(u_{1_1}(1-T_1)
+ (1-u_{1_1})(1-T_1)[u_{2_1}(1-T_2) + (1-u_{2_1})\bar f_3(u_{3_1}) + u_{2_1}T_2\bar f_3(u_{3_2})]\nonumber\\
& + u_{1_1}T_1[u_{2_2}(1-T_2) + (1-u_{2_2})\bar f_3(u_{3_3}) + u_{2_2}T_2\bar f_3(u_{3_4})]) 
    - \frac{1}{u_{1_1}(1-u_{2_2})} G_1(u_{1_1}(1-T_1) \nonumber\\
    &+ (1-u_{1_1})(1-T_1)[u_{2_1}(1-T_2) + (1-u_{2_1})(1-T_2)\bar f_3(u_{3_1}) + u_{2_1}T_2\bar f_3(u_{3_2})]\nonumber\\&
+ u_{1_1}T_1[u_{2_2}(1-T_2) + (1-u_{2_2})(1-T_2)\bar f_3(u_{3_3}) + u_{2_2}T_2\bar f_3(u_{3_4})]) 
\\ \nonumber \\
    u_{3_4}=& \frac{1}{u_{1_1}u_{2_2}} G_1(u_{1_1}(1-T_1)+ (1-u_{1_1})(1-T_1)[u_{2_1}(1-T_2) + (1-u_{2_1})(1-T_2)\bar f_3(u_{3_1}) + u_{2_1}T_2\bar f_3(u_{3_2})]\nonumber\\
    &+ u_{1_1}T_1[u_{2_2}(1-T_2) + (1-u_{2_2})(1-T_2)\bar f_3(u_{3_3}) + u_{2_2}T_2\bar f_3(u_{3_4})]) 
\end{alignat}
\end{subequations}

The outbreak size of the third strain is then given by 

\begin{align}
        \mathcal A_3 =& \mathcal A_2-\bigg[G_0(u_{1_1}(1-T_1)+ (1-u_{1_1})[u_{2_1}(1-T_2) + (1-u_{2_1})\bar f_3(u_{3_1}) + u_{2_1}T_2\bar f_3(u_{3_2})]\nonumber\\
    &+ u_{1_1}T_1[u_{2_2}(1-T_2) + (1-u_{2_2})\bar f_3(u_{3_3}) + u_{2_2}T_2\bar f_3(u_{3_4})]) \nonumber\\
    &- 
    G_0(u_{1_1}(1-T_1)+ (1-u_{1_1})[u_{2_1}(1-T_2) + (1-u_{2_1})(1-T_2)\bar f_3(u_{3_1}) + u_{2_1}T_2\bar f_3(u_{3_2})]\nonumber\\
    &+ u_{1_1}T_1[u_{2_2}(1-T_2) + (1-u_{2_2})(1-T_2)\bar f_3(u_{3_3}) + u_{2_2}T_2\bar f_3(u_{3_4})]) \nonumber\\
    &- 
   G_0(u_{1_1}(1-T_1)+ (1-u_{1_1})(1-T_1)[u_{2_1}(1-T_2) + (1-u_{2_1})\bar f_3(u_{3_1}) + u_{2_1}T_2\bar f_3(u_{3_2})]\nonumber\\
    &+ u_{1_1}T_1[u_{2_2}(1-T_2) + (1-u_{2_2})\bar f_3(u_{3_3}) + u_{2_2}T_2\bar f_3(u_{3_4})]) 
    \nonumber\\
    &+ 
    G_0(u_{1_1}(1-T_1)+ (1-u_{1_1})(1-T_1)[u_{2_1}(1-T_2) + (1-u_{2_1})(1-T_2)\bar f_3(u_{3_1}) + u_{2_1}T_2\bar f_3(u_{3_2})]\nonumber\\
    &+ u_{1_1}T_1[u_{2_2}(1-T_2) + (1-u_{2_2})(1-T_2)\bar f_3(u_{3_3}) + u_{2_2}T_2\bar f_3(u_{3_4})])  \bigg]
\end{align}
which is simply $u_{3_1}$, the maximally externally coinfected state, without the prior probabilities and with $G_1(z)$ replaced by $G_0(z)$.

\subsection*{Strain 4}

In this section we evaluate the expressions for the outbreak size of strain 4 for collaborative percolation. The first step is to find the base probability. To do this, we require $\mathcal C_4=\bar{\mathcal C}_3(\bar f_4)$, which is quite a complicated nested function. However, figures \ref{fig:strain4NEST} and \ref{fig:iterated_type2strain4NEST} show how this expression is formed graphically as a perfect binary tree. 

\begin{figure}[ht!]
  \begin{center}
    \includegraphics[width=0.65\textwidth]{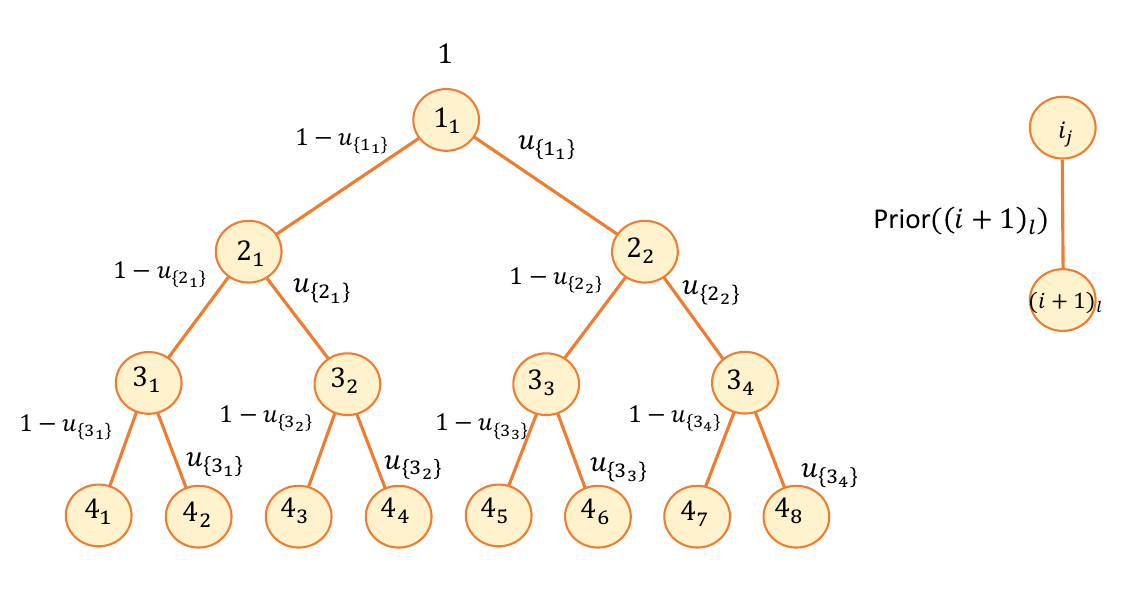}
  \end{center}
  \caption{A visualisation of the perfect binary tree of coinfected neighbours which a focal node could be surrounded by after strain 3. The states are marked inside the nodes, whilst the priors can be constructed by the product of the terms associated to each edge from the root to each leaf. External infection occurs on the left child, whilst direct infection occurs on the right.
  }
  \label{fig:strain4NEST}
\end{figure}

\begin{figure}[ht!]
  \begin{center}
    \includegraphics[width=0.99\textwidth]{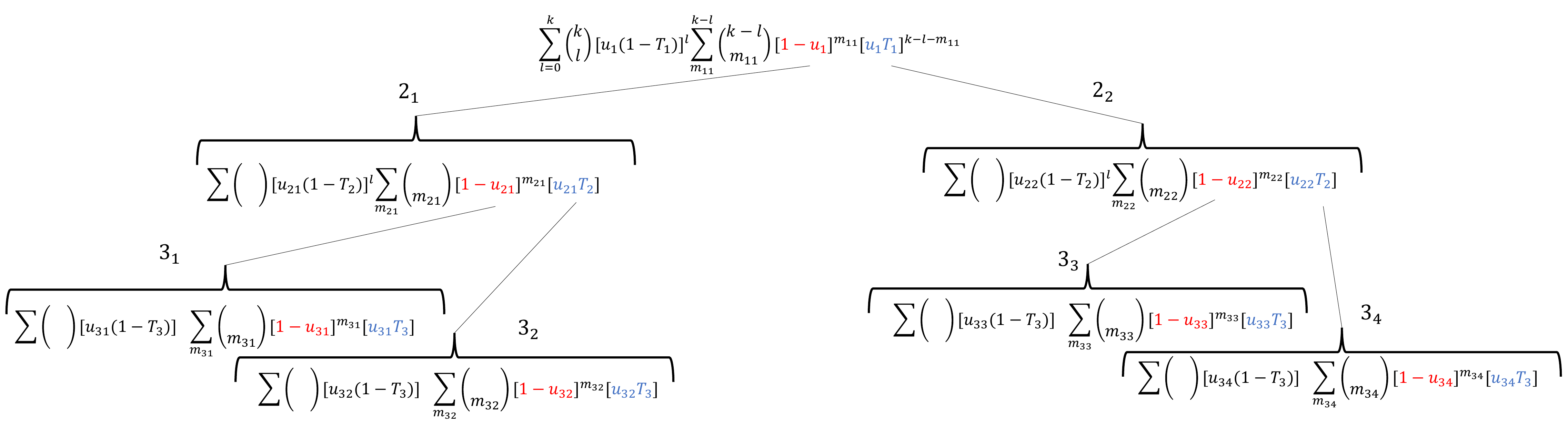}
  \end{center}
  \caption{A visualisation of the perfect binary tree of coinfected neighbours which a focal node could be surrounded by and the equations that generate the base probability $\bar{\mathcal C}_3(\bm 1)$. In this notation, the left child always represents external infection whilst the right child represents direct infection, as in Fig \ref{fig:strain4NEST}. Thus, the leaves of the tree represent all possible infection histories for the 4th strain. 
  }
  \label{fig:iterated_type2strain4NEST}
\end{figure}

\begin{align}
    \bar{\mathcal C}_3(\bm 1) =&\  \sum_{l=0}^k\binom{k}{l}[u_{1_1}(1-T_1)]^l\nonumber\\
    &\ \times\sum_{m_{1_1}=0}^{k-l}\binom{k-l}{m_{1_1}}[1-u_{1_1}]^{m_{1_1}}\sum_{a=0}^{m_{1_1}}\binom{m_{1_1}}{a}[u_{2_1}(1-T_2)]^a\sum_{m_{2_1}=0}^{m_{1_1}-a}\binom{m_{1_1}-a}{m_{2_1}}[(1-u_{2_1})]^{m_{2_1}}\nonumber\\
    &\ \times \sum^{m_{2_1}}_{b=0} \binom{m_{2_1}}{b} [u_{3_1}(1-T_3)]^{b}\sum^{m_{2_1}-b}_{m_{3_1}=0}\binom{m_{2_1}-b}{m_{3_1}} [1-u_{3_1}]^{m_{3_1}}[u_{3_1}T_3]^{m_{2_1}-b-m_{3_1}}[u_{2_1}T_2]^{m_{1_1}-a-m_{2_1}}
    \nonumber\\
    &\ \times 
    \sum^{m_{1_1}-a-m_{2_1}}_{c=0} \binom{m_{1_1}-a-m_{2_1}}{c} [u_{3_2}(1-T_3)]^{c}\sum^{m_{1_1}-a-m_{2_1}-c}_{m_{3_2}=0}\binom{m_{1_1}-a-m_{2_1}-c}{m_{3_2}} [1-u_{3_2}]^{m_{3_2}}[u_{3_2}T_3]^{m_{1_1}-a-m_{2_1}-c-m_{3_2}}\nonumber\\
    &\ \times [u_{1_1}T_1]^{k-l-m_{1_1}}\sum_{e=0}^{k-l-m_{1_1}}\binom {k-l-m_{1_1}}{e}[u_{2_2}(1-T_2)]^e\sum_{m_{2_2}=0}^{k-l-m_{1_1}-e}\binom{k-l-m_{1_1}-e}{m_{2_2}}[1-u_{2_2}]^{m_{2_2}}
    \nonumber\\
    &\ \times
    \sum^{m_{2_2}}_{f=0} \binom{m_{2_2}}{f} [u_{3_3}(1-T_3)]^{f}\sum_{m_{33}=0}^{m_{2_2}-f}\binom{m_{2_2}-f}{m_{3_3}} [1-u_{3_3}]^{m_{3_3}}[u_{3_3}T_3]^{m_{2_2}-f-m_{3_3}}[u_{2_2}T_2]^{k-l-m_{1_1}-e-m_{2_2}}\nonumber\\
    &\ \times
    \sum^{k-l-m_{1_1}-e-m_{2_2}}_{g=0} \binom{k-l-m_{1_1}-e-m_{2_2}}{g} [u_{3_4}(1-T_3)]^{g}\sum_{m_{3_4}=0}^{k-l-m_{1_1}-e-m_{2_2}-g}\binom{k-l-m_{1_1}-e-m_{2_2}-g}{m_{3_4}}\nonumber\\
    &\ \times [1-u_{3_4}]^{m_{3_4}}[u_{3_4}T_3]^{k-l-m_{1_1}-e-m_{2_2}-g-m_{3_4}}
\end{align}
We then have
\begin{align}
     {\mathcal C_4}=&\ f_1(u_1,f_2(u_{2_1},f_3(u_{3_1},\bar f_4(u_{4_1}) ,\bar f_4(u_{4_2}) ),f_3(u_{3_2},\bar f_4(u_{4_3}) ,\bar f_4(u_{4_4}) )),\nonumber\\
     &\times f_2(u_{2_2},f_3(u_{3_3},\bar f_4(u_{4_5}) ,\bar f_4(u_{4_6}) ),f_3(u_{3_4},\bar f_4(u_{4_7}) ,\bar f_4(u_{4_8}))))
\end{align}
The next step is to formulate the 8 $u_{4_i}$ values given each unique infection history. These are given by 
\begin{subequations}
\begin{align}
    u_{4_1} =\ &\frac{1}{(1-u_{1_1})(1-u_{2_1})(1-u_{3_1})}\sum^\infty_{k=0}q_k \mathcal C_4\mathcal H_{3_1}[1-(1-T_3)^{m_{3_1}+m_{3_2}+m_{3_3}+m_{3_4}}]\\
    u_{4_2} =\ &\frac{1}{(1-u_{1_1})(1-u_{2_1})u_{3_1}}\sum^\infty_{k=0}q_k \mathcal C_4\mathcal H_{3_1}(1-T_3)^{m_{3_1}+m_{3_2}+m_{3_3}+m_{3_4}}\\
    u_{4_3} =\ & \frac{1}{(1-u_{1_1})u_{2_1}(1-u_{3_2})} \sum^\infty_{k=0}q_k \mathcal C_4\mathcal H_{3_2}[1-(1-T_3)^{m_{3_1}+m_{3_2}+m_{3_3}+m_{3_4}}]\\
    u_{4_4} =\ & \frac{1}{(1-u_{1_1})u_{2_1}u_{3_2}}\sum^\infty_{k=0}q_k \mathcal C_4\mathcal H_{3_2}(1-T_3)^{m_{3_1}+m_{3_2}+m_{3_3}+m_{3_4}}\\
    u_{4_5} =\ & \frac{1}{u_{1_1}(1-u_{2_2})(1-u_{3_3})}\sum^\infty_{k=0}q_k \mathcal C_4\mathcal H_{3_3}[1-(1-T_3)^{m_{3_1}+m_{3_2}+m_{3_3}+m_{3_4}}]\\
    u_{4_6} =\ & \frac{1}{u_{1_1}(1-u_{2_2})u_{3_3}}\sum^\infty_{k=0}q_k \mathcal C_4\mathcal H_{3_3}(1-T_3)^{m_{3_1}+m_{3_2}+m_{3_3}+m_{3_4}}\\
    u_{4_7} =\ & \frac{1}{u_{1_1}u_{2_2}(1-u_{3_4})}\sum^\infty_{k=0}q_k \mathcal C_4\mathcal H_{3_4}[1-(1-T_3)^{m_{3_1}+m_{3_2}+m_{3_3}+m_{3_4}}]\\
    u_{4_8} =\ & \frac{1}{u_{1_1}u_{2_2}u_{3_4}}\sum^\infty_{k=0}q_k \mathcal C_4\mathcal H_{3_4}(1-T_3)^{m_{3_1}+m_{3_2}+m_{3_3}+m_{3_4}}\\
\end{align}
\end{subequations}
We show the solution to these expressions graphically in Fig \ref{fig:non-trivialroot4} around the critical point and for large $T$ values. The outbreak size is then given by 
\begin{equation}
    \mathcal A_4 = \mathcal A_3 - \sum^\infty_{k=0}p_k \mathcal C_4\mathcal H_{3_1}[1-(1-T_3)^{m_{3_1}+m_{3_2}+m_{3_3}+m_{3_4}}]
\end{equation}

\begin{figure}[ht!]
  \begin{center}
    \includegraphics[width=0.85\textwidth]{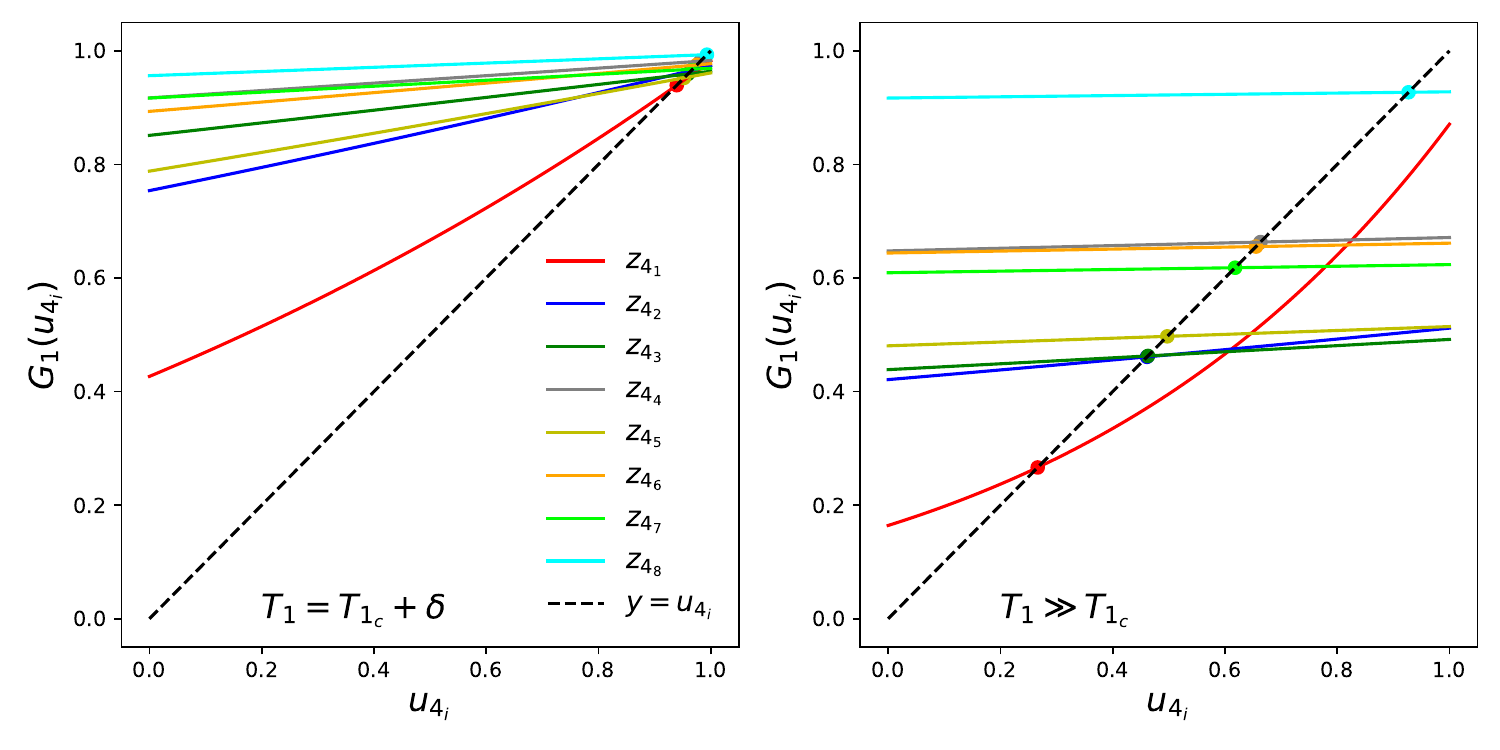}
  \end{center}
  \caption{The graphical solution of the generating functions for strain 4 around the critical point of strain 1 (left) with $T_1=0.325,T_2=0.6,T_3 = 0.5,T_4 = 0.45$ and away from the critical point at $T_1=T_2=0.6,T_3 = 0.5,
T_4 = 0.45$ (right). Plotted are $y=u_{4_i}$ for $i=1,\dots,8$ against $u_{4_i}$ as well as the value of the generating functions $z_i=G_1(\mathcal C_2\mathcal H_{4_i})/\mathcal Q_{4_i}$. Each $z_i$ varies $u_{4_i}$ whilst the other value held fixed at the correct root. The intersection of $y=u_{4_i}$ and $z_i$ corresponds to the root, which is also marked with a scatter point from a non-linear solve. We notice that $u_{4_1}$ varies convexly over almost the entire unit interval whilst the gradient of the other generating functions is increasingly flat in this region of the parameter space. This indicates that the contribution of these values is less important than that $u_{4_1}$. Thus, we can expect that this infection history is the dominant term in the non-linear system that describes strain 4. 
  }
  \label{fig:non-trivialroot4}
\end{figure}

\subsection*{Strain 5}

Following the recipe, the outbreak size of strain 5 is calculated as follows. 

\begin{align}
    \mathcal C_5 = &\ f_1(u_1,f_2(u_{2_1},f_3(u_{3_1}, f_4(u_{4_1},\bar f_5(u_{5_1}),\bar f_5(u_{5_2})) , f_4(u_{4_2},\bar f_5(u_{5_3}),\bar f_5(u_{5_4})) ),\nonumber\\
     &\times f_3(u_{3_2}, f_4(u_{4_3},\bar f_5(u_{5_5}),\bar f_5(u_{5_6})) , f_4(u_{4_4},\bar f_5(u_{5_7}),\bar f_5(u_{5_8})) )),\nonumber\\
     &\times f_2(u_{2_2},f_3(u_{3_3}, f_4(u_{4_5},\bar f_5(u_{5_9}),\bar f_5(u_{5_{10}})) , f_4(u_{4_6},\bar f_5(u_{5_{11}}),\bar f_5(u_{5_{12}})) ),\nonumber\\
     &\times f_3(u_{3_4}, f_4(u_{4_7},\bar f_5(u_{5_{13}}),\bar f_5(u_{5_{14}})) , f_4(u_{4_8},\bar f_5(u_{5_{15}}),\bar f_5(u_{5_{16}})))))
\end{align}

\begin{subequations}
\begin{align}
    u_{5_{1}} =\ & \frac{1}{(1-u_{1_1})(1-u_{2_1})(1-u_{3_1})(1-u_{4_1})}\sum^\infty_{k=0}q_k\mathcal C_5 H_{4_{1}}[1-(1-T_4)^{M}]  \\
     u_{5_{2}} =\ & \frac{1}{(1-u_{1_1})(1-u_{2_1})(1-u_{3_1})u_{4_1}}\sum^\infty_{k=0}q_k\mathcal C_5 H_{4_{1}}(1-T_4)^{M}  \\
      u_{5_{3}} =\ & \frac{1}{(1-u_{1_1})(1-u_{2_1})u_{3_1}(1-u_{4_2})}\sum^\infty_{k=0}q_k\mathcal C_5 H_{4_{2}} [1-(1-T_4)^{M}]  \\
       u_{5_{4}} =\ & \frac{1}{{(1-u_{1_1})(1-u_{2_1})u_{3_1}u_{4_2}}}\sum^\infty_{k=0}q_k\mathcal C_5 H_{4_{2}}(1-T_4)^{M}   \\
        u_{5_{5}} =\ & \frac{1}{(1-u_{1_1})u_{2_1}(1-u_{3_2})(1-u_{4_3})}\sum^\infty_{k=0}q_k\mathcal C_5 H_{4_{3}}[1-(1-T_4)^{M}]   \\
         u_{5_{6}} =\ & \frac{1}{(1-u_{1_1})u_{2_1}(1-u_{3_2})u_{4_3}}\sum^\infty_{k=0}q_k\mathcal C_5 H_{4_{3}} (1-T_4)^{M}  \\
          u_{5_{7}} =\ & \frac{1}{(1-u_{1_1})u_{2_1}u_{3_2}(1-u_{4_4})}\sum^\infty_{k=0}q_k\mathcal C_5 H_{4_{4}}[1-(1-T_4)^{M}]   \\
           u_{5_{8}} =\ & \frac{1}{(1-u_{1_1})u_{2_1}u_{3_2}u_{4_4}}\sum^\infty_{k=0}q_k\mathcal C_5 H_{4_{4}}(1-T_4)^{M}   \\
            u_{5_{9}} =\ & \frac{1}{u_{1_1}(1-u_{2_2})(1-u_{3_3})(1-u_{4_5})}\sum^\infty_{k=0}q_k\mathcal C_5 H_{4_{5}} [1-(1-T_4)^{M}]  \\
             u_{5_{10}} =\ & \frac{1}{u_{1_1}(1-u_{2_2})(1-u_{3_3})u_{4_5}}\sum^\infty_{k=0}q_k\mathcal C_5 H_{4_{5}}(1-T_4)^{M}   \\
              u_{5_{11}} =\ & \frac{1}{u_{1_1}(1-u_{2_2})u_{3_3}(1-u_{4_6})}\sum^\infty_{k=0}q_k\mathcal C_5 H_{4_{6}}[1-(1-T_4)^{M}]   \\
               u_{5_{12}} =\ & \frac{1}{u_{1_1}(1-u_{2_2})u_{3_3}u_{4_6}}\sum^\infty_{k=0}q_k\mathcal C_5 H_{4_{6}}(1-T_4)^{M}   \\
                u_{5_{13}} =\ & \frac{1}{u_{1_1}u_{2_2}(1-u_{3_4})(1-u_{4_7})}\sum^\infty_{k=0}q_k\mathcal C_5 H_{4_{7}} [1-(1-T_4)^{M}]  \\
                 u_{5_{14}} =\ & \frac{1}{u_{1_1}u_{2_2}(1-u_{3_4})u_{4_7}}\sum^\infty_{k=0}q_k\mathcal C_5 H_{4_{7}}(1-T_4)^{M}   \\
                  u_{5_{15}} =\ & \frac{1}{u_{1_1}u_{2_2}u_{3_4}(1-u_{4_8})}\sum^\infty_{k=0}q_k\mathcal C_5 H_{4_{8}} [1-(1-T_4)^{M}]  \\
                   u_{5_{16}} =\ & \frac{1}{u_{1_1}u_{2_2}u_{3_4}u_{4_8}}\sum^\infty_{k=0}q_k\mathcal C_5 H_{4_{8}}(1-T_4)^{M}   \\
\end{align}
\end{subequations}
with $M = \sum\limits_{j=1}^8 m_{4_j}$. The outbreak size is then given by 
\begin{equation}
     \mathcal A_5 =\mathcal A_4 - \sum^\infty_{k=0}p_k\mathcal C_5 H_{4_{1}}[1-(1-T_4)^{M}] 
\end{equation}

\end{widetext}

\end{document}